\documentstyle[11pt]{article}
\begin{document}
\def \xa { (1+ 2 \nu_\alpha )^2  }
\def \xb { (1+ 2 \nu_\beta )^2  }
\setlength{\baselineskip}{0.6cm}

\begin{center}
{\LARGE {\bf Quantum  Corrections to the Classical Dynamics: \\
Application to the SU(2) Lipkin Model}} 
\footnote{\it This work was partially supported by FAPESP and
 CNPq.}
\end{center}
\vspace{0.7cm}

\begin{center}
{\large {\bf M.Trindade dos Santos$^{(1)}$, M.C.Nemes$^{(2)}$}}
\end{center}

\vspace{0.5cm}

{\it {\large $^{(1)}$ Universidade de S\~ao Paulo,

Instituto de F\'\i sica, 

Departamento de F\'\i sica Matem\'atica,

C.P. 66318 05389-970 S\~ao Paulo, SP, Brazil.

E-mail: msantos@if.usp.br
\vspace{0.5cm} 

$^{(2)}$ Universidade Federal de Minas Gerais,

Departamento de F\'\i sica, ICEX,

C.P. 702, 30161-970 Belo Horizonte, MG, Brazil.

E-mail: carolina@fisica.ufmg.br
}}

\vspace{1.0cm}

\begin{abstract}
We show how nonrelativistic many body techniques can be used to study 
quantum corrections to the classical limit, in particular of the
$SU(2)$ Lipkin Model. We show that the quantum corrections are essentially
of two types: unitary and nonunitary. In this work we perform a detailed study
of the unitary corrections. They can be cast in Hamiltonian form and are 
shown to double
the number of degrees of freedom. As a consequence chaotic behavior emerges.
We show that this semiquantal chaos is the mechanism trough which tunneling
is effected. We also show that these corrections systematically improve the
classical results and propose some quantitative measure of this improvement. 

\noindent
PAC number (21.10)
\end{abstract}

\newpage
\section{Introduction}

\vspace{1.0cm}

One of the most fascinating open problems in the area of dynamical systems is
the search for an adequate semiclassical description of quantum systems.
Several attempts focus on obtaining $\hbar$ corrections to the dynamics of 
the corresponding classical system. A pioneer work along this idea is the
semiclassical method proposed by Einstein-Brilloin-Keller \cite{Percival}
which leads to quantization rules for integrable systems. Other 
important contributions are approximations to the Feynman path integral
formulation \cite{Schulman}, used to derive the periodic orbit trace formula
for chaotic systems: The Gutzwiller trace formula \cite{Gutzwiller}. 
This relates the spectrum of
 quantum systems to a weighted sum over the unstable periodic orbits of 
the classical system.

Recently semiquantal dynamics has been derived via Ehrenfest's theorem
and recast as an extended classical system with the fluctuation variables
 coupled to the average variables \cite{Pattanyak1} \cite{Pattanyak2}.
 A different 
approach which can be shown to yield identical results is the time dependent
 variational principle where the true solution is approximated by 
squeezed states, the so called gaussian variational approximation
\cite{Jackiw}, \cite{Cooper}, \cite{Kovner} and \cite{Lin1}.
Other methods for this propose are quoted in ref. \cite{Eckhardt}.
The Wigner formalism has also recently been applied to study diffusive
and dissipative type of corrections \cite{Italianos1} and \cite{Italianos2} .

The purpose of the present contribution is twofold. The first one is of
formal character. We show that a mean field expansion in the sense of 
nonrelativistic many body theories can be used as  a consistent and 
systematic tool to
analyse the nature of quantum corrections to the classical limit.
We show that such corrections can be classified in two types:
unitary quantum corrections, which amount to considering
(on top of the dynamics of the average values) the dynamics of the width
of gaussian wave packets. The number of degrees of freedom of this system is
therefore doubled, and the semiclassical limit can be cast into the form of
Hamilton's equations. This unitary dynamics reveals in particular general
features of quantum kinematics - a centrifugal type barrier involving
the width of the packet is the classical counterpart of the uncertainty 
principle - and the resulting dynamics is in general chaotic even if we start 
from a simple integrable system, as we will show. The same result in spirit
has been obtained in the context of a variational calculation by Pattanyak 
and Schieve \cite{Pattanyak1} in the framework of Heisenberg's equations
of motion. The second type of corrections to the classical limit in the
 context of many body mean field calculation is non unitary in character.
It arises from the inclusion of quantum correlation contributions.
 Their inclusion induce, given adequate approximations,
 a Langevin-type force on top of the 
Hamiltonian description provided by the unitary evolution. They can thus
formally account for diffusive and dissipative type behaviors, making 
immediate contact with the work of reference \cite{Italianos1}.

The second purpose of the present work is to explore in detail the unitary
type corrections in the context of the integrable $SU(2)$ Lipkin Model 
\cite{Lipkin}.
The reason for choosing this particular model, aside  its simplicity,
resides in the fact that it possesses a well defined classical limit  when the 
number of particles $N$ goes to infinity \cite{Yaffe}. In this case it can
 be rigorously
shown that the classical limit corresponds to constructing the Hamiltonian 
dynamics for the parameters of coherent states, a special case of gaussian
wave packets. In this case we will be safe to perform comparisons between 
the classical, unitary semiclassical and exact dynamics. We show that the
derived corrections improve the classical results both qualitatively and
quantitatively. 
From the qualitative point of view we show that the presence of quantum
degrees of freedom (widths of wave packet) induces chaotic motion and
this is the mechanism through which quantum tunneling is effected, a clear
quantum correction to the classical description. Also from the qualitative
point of view we analytically study the time evolution of observable for
short times, investigating in this way modifications introduced by quantum 
corrections. 
From the quantitative point of view, in order to assess the time of validity
of the approximation we perform a comparative numerical study
of the time evolution given by the approximations and the exact result for
the time evolution of observable. We show that despite of chaotic behavior
the semiclassical or gaussian approximation gives a better description of
 the dynamics. We also set up a quantitative measure of the time of validity
of the approximation according to which the gaussian approximation is 
better than the classical one and this quality increases with $N$, as it
should.

This paper is divided as follows: section $2$ contains the formalism based 
on which we define the semiclassical limit, section $3$ contains an
 application to the $SU(2)$ Lipkin Model, section $4$ contains a discussion of
our numerical results and section $5$ contains some conclusions.

\newpage

\section{The Formalism:}

One of the most widely used method to construct the classical limit of a 
quantum system is by means of coherent states \cite{Perelomov}.
 As is well known such method
can be viewed as a mean field approximation where the width of the wave 
packet is minimal and remains unchanged during the time evolution.

The idea of defining the semiclassical limit as  some kind of mean field
approximation is appealing. A very general and unique definition of 
mean field approximation can be given once one requires that the density
function  which should be used to calculate traces be the one
which reproduces in exact form all expectation values of 
one body operators. In other words a gaussian density operator \cite{Qnoise}.

The formalism stated below is a simple application of techniques developed
before for the treatment of the reduced dynamics of gaussian observable of 
interacting many boson systems in Many Body Nuclear Physics and 
Quantum Field Theory \cite{Lin2}. We make here an option for 
self-containedness.
In the eventual lack of technical details the reader is referred
 to the works in references \cite{Lin1}, \cite{Lin2},  and \cite{squeezing}.

Quantum bosonic states are represented by density operators $F$ so that
mean values of a chosen  operator ${\cal{O}}$ are given in terms 
of traces, e.g. 

\begin{equation}
<  {\cal{O}}> =Tr({\cal{O}} F) \;\;\;,
\end{equation}

\noindent
and the basic dynamical equation is the Liouville-von Neumann equation for $F$
\begin{equation}
\imath \dot{F}=[H,F] \;\;\; ,
\end{equation}

\noindent
where $H$ is the Hamiltonian of the system.

Gaussian states are particular densities which are fully determined by the
mean values  of the field operators and their bilinear or quadratic 
combinations, in addition to statistics. For bosonic systems 
($[a,a^{\dag}]=1$) the relevant
quantities are $ <a^{\dag}>$ , $< a^{\dag} a>$ , and $< a^{\dag} a^{\dag}>$.
The first of these can be conveniently parameterized in terms of two real
quantities $q$ and $p$

\begin{equation}
< a^{\dag}>=Tr(a^{\dag}F_0)=\sqrt{\frac{\mu_0}{2}}
(q- \imath \frac{p}{\mu_0})
\;\;\; ,
\label{eq3}
\end{equation}

\noindent
where $F_0$ is the density matrix (it will be defined later) and $\mu_0$ is a
scale parameter (we set $\mu_0=1$).
To deal with the other two one can define new Bogoliubov quasiboson 
operators as 
 
\begin{equation}
\left(
\begin{array}{c}
\tilde \eta \\
\tilde\eta^{\dag}
\end{array}
\right) 
=\bf{X}^{\dag} 
\left(
\begin{array}{c}
\tilde a\\
\tilde a^{\dag}
\end{array}
\right) \;\;\; ,
\end{equation}

\noindent
where $\tilde \eta=\eta -<\eta>$, $\tilde a = a - <a>$ and  $\bf{X}$ define 
the Bogoliubov transformation. It is given by 

\begin{equation}
\bf{X}=
\left(
\begin{array}{cc}
x^* & y \\
y^* & x
\end{array}
\right) \;\;\; ,
\end{equation}

\noindent
satisfying the normalization condition

\begin{equation}
\bf{X^{\dag} G X=G} 
\end{equation}

\noindent
with
\begin{equation}
\bf{G}=
\left(
\begin{array}{cc}
1 & 0 \\
0 & -1
\end{array}
\right) \hspace{1.5cm} , \hspace{1.5cm} \bf{G}^2 = \bf{1} \;\;\;.
\end{equation}

The preservation of the commutation relations $[\eta, \eta^{\dag}]=1 $
 requires as usual  that the 
transformation coefficients $x$ and $y$ be chosen so that $|x|^2 - |y|^2 =1$.
This is guaranteed by equation (6).

For the bilinear operators equation (4) implies that

\begin{equation}
\bf{N= X^{\dag} R X} \;\;\;,
\end{equation}

\noindent
where
\begin{equation}
\bf{N}=
\left(
\begin{array}{cc}
<\tilde\eta^{\dag} \tilde\eta> & <\tilde\eta \tilde\eta>  \\
<\tilde\eta^{\dag} \tilde\eta^{\dag}>  & <\tilde\eta \tilde\eta^{\dag}>
\end{array}
\right)
\;\;\;\;\; ,  \;\;\;\;\;
\bf{R}=
\left(
\begin{array}{cc}
<\tilde a^{\dag} \tilde a> &< \tilde a \tilde a>\\
<\tilde a^{\dag} \tilde a^{\dag} > & <\tilde a \tilde a^{\dag}>
\end{array}
\right)\;\;.
\end{equation}

This Bogoliubov transformation is so chosen that from a direct calculation
one has

\begin{equation}
Tr(\tilde{\eta}\tilde{\eta}F_0)
=Tr(\tilde{\eta}^{\dag}\tilde{\eta}^{\dag}F_0)=0
\;\;\; .
\end{equation}

\noindent
This gives us also the quantity  
$\nu=<\tilde{\eta}^{\dag} \tilde{\eta}>$ which
is the occupation probability. We can thus parameterize $<a^{\dag}a^{\dag}>$
and $<a^{\dag} a>$ in terms of $x$, $y$ and $\nu$. As a consequence of this
the dispersion of the quadratures $\Delta q$ and $\Delta p$ must depend on 
x,y and $\nu$ only

\[
\Delta q^2 = \frac{1}{2 \mu_0}[ <(a^{\dag} + a)> ^2 -
( < a>^* + <a>)^2]=
\]

\begin{equation}
= \frac{1}{2 \mu_0}[1 + 2|y|^2 -(x^*y+y^*x)](1+2 \nu) \;\;\;,
\end{equation}

\[
\Delta p^2 = - 2 \mu_0 [ <(a^{\dag} -a)^2 > -
( < a >^* - < a  > )^2]=
\]

\begin{equation}
= 2 \mu_0[1 + 2|y|^2 +(x^*y+y^*x)](1+2 \nu) \;\;\; .
\end{equation}

As we will show in the sequel it is possible to define a pair of canonically
conjugate variables associated to $\Delta q$ and $\Delta p$ as follows

\[
\Delta q^2=Q^2\;\;\;,
\]

\[
\Delta p^2 = P^2 + \frac{(2 \nu +1)}{Q^2} \;\;\;.
\]

 The gaussian state so specified is given by

\begin{equation}
F_0=\frac{1}{1+\nu}
(\frac{\nu}{1+\nu})^{\tilde{\eta}^{\dag} \tilde{\eta}} 
\end{equation}
\[
=\sum_{n} |n> \frac{1}{1+\nu} (\frac{\nu}{1+\nu}) ^n <n| \;\;\;,
\]

\noindent
where $\tilde{\eta}^{\dag} \tilde{\eta} |n>= \nu |n>$, is the state which
contains all the information about the operators in question and only this.
Notice that the operator $\tilde{\eta}^{\dag} \tilde{\eta}$ in the expression
of $F_0$ is a linear combination of the bosonic operators
$ a^{\dag} , a , a^{\dag} a, a a , a^{\dag} a^{\dag} $ and the coefficients
are functions of $q,p,x,y$ and $\nu$. 
More technical details are given in the didactic work by de Cloizeaux
\cite{Cloizeaux}. 
It is a simple matter to check that $F_0$ need not be a pure state and that 
in the limit $\nu \rightarrow 0$ it goes to the pure gaussian state
$|0><0|$. This state precisely corresponds  to the variational state
 used in refs. \cite{Jackiw} and \cite{Kovner}.
 The usual coherent state approximation corresponds to 
setting  $\nu=0$ and moreover requiring the fullfilment of the minimal 
uncertainty relation, $x=1$ and $y=0$ for all times.

Notice that the equation 
$Tr( \tilde{\eta} F_0) =  Tr( \tilde{\eta}^{\dag}  F_0)=0$ is a consequence of
the definition of $\tilde{\eta}$ and therefore imposes no constraints on
$x$ and $y$. Therefore we have four undetermined parameters Re$(x)$ ,Im$(x)$,
Re$(y)$ and Im$(y)$ and three conditions to fix them, namely, eq. (10) and 
the normalization condition. The fourth one can be obtained by imposing an 
overall null phase in the state $F_0$.

We now sketch the mean field expansion method: In general the state 
$F$ is not of the form $F_0$, but even so can be used to evaluate 
$<a^{\dag}>, <a^{\dag}a^{\dag}>$ and $<a^{\dag}a>$ and hence a set of gaussian
parameters. In terms of these we can set up a gaussian state $F_0$ and
split the complete state $F$ as

\begin{equation}
F=F_0+F'
\end{equation}

\noindent
where $Tr(F')=0$, so that $F'$ is a pure correlation part of $F$. We
make essential use of the fact that the decomposition (14) can be implemented
in terms of  a projection operator $P(t)$ such that $F_0=P(t)F$ 
(the explicit form of $P(t)$ for bosons is given in refs. 
\cite{Lin1} and \cite{Lin2})
to rewrite eq. (14) as a set of coupled equations for $F_0$ and $F'(t)$.
This eventually allows one to write $F'(t)$ as a function of the past 
history of the gaussian projection $F_0(t)$:

\begin{equation}
F'(t)=F'[ F_0(t'<t)]
\end{equation}

Using this result one can close the equations of motion for the gaussian 
parameters again by taking appropriate traces of the Liouville - von Neuman 
equation (2) and using $F'(t)=F_0(t) + F'[ F_0(t'<t)]$.
The first term will reproduce the mean field result, while the second will
 give rise to additional terms involving memory integrals 
(refs \cite{Lin1} and \cite{Lin2}).
As a matter of fact, the expression for equation (15) is in general not 
computable even in simple model problems without approximations. However
a consistent and systematic approximation scheme has been constructed based 
on a criterium of energy conservation \cite{Feld}

\[
<H>=Tr(H F_0) + Tr(H F')
\]

\noindent
so that to every order of approximation one makes sure that $d<H>/dt=0$.
As shown in ref. \cite{Feld} and implemented in various systems 
\cite{squeezing} \cite{Erica} this 
criterium leads to a systematic, controllable expansion around the mean
field approximation.

Since we have defined $\nu=Tr(F_0 \tilde\eta^{\dag} \tilde\eta)$ the occupation
probability does not evolve on time on the mean field approximation level
(unitary contributions).
We have

\[
\imath \dot\nu = Tr[\tilde{\eta}^{\dag} \tilde{\eta},H] F_0=
Tr H [F_0, \tilde{\eta}^{\dag} \tilde{\eta}] \;\;\;, 
\]

\noindent 
and $[F_0,\tilde{\eta}^{\dag} \tilde{\eta}] \equiv 0$ (see eq.(13)) 
The nonunitary contributions will come from the time evolution of quantum 
correlations, 

\begin{equation}
\imath \frac{d}{dt} \tilde{\eta}^{\dag} \tilde{\eta}= \imath\dot{\nu}(t)=
Tr ([\tilde{\eta}^{\dag} \tilde{\eta},H] F') \;\;\;,
\end{equation}

\noindent 
in the form of an explicit time dependence on the occupation probability
$\nu$. This has been explicitly
 studied in the context of the anharmonic oscillator \cite{Lin2}
 and of the Maser Model \cite{Erica},
 showing both qualitatively and quantitatively what are the 
effects of many body correlations at the level of one body observables.
In the present work these nonunitary contributions have been neglected,
since they are of higher order than the unitary ones.
It is however worthwhile noticing that the inclusion of such corrections may 
lead to Fokker-Planck type equations making thus immediate contact with the
works of refs. \cite{Italianos1} and \cite{Italianos2}.

Let us now make the connection between this general mean field expansion
and the classical limit plus corrections. From the point of view of this
formalism the classical limit corresponds to the following scheme:

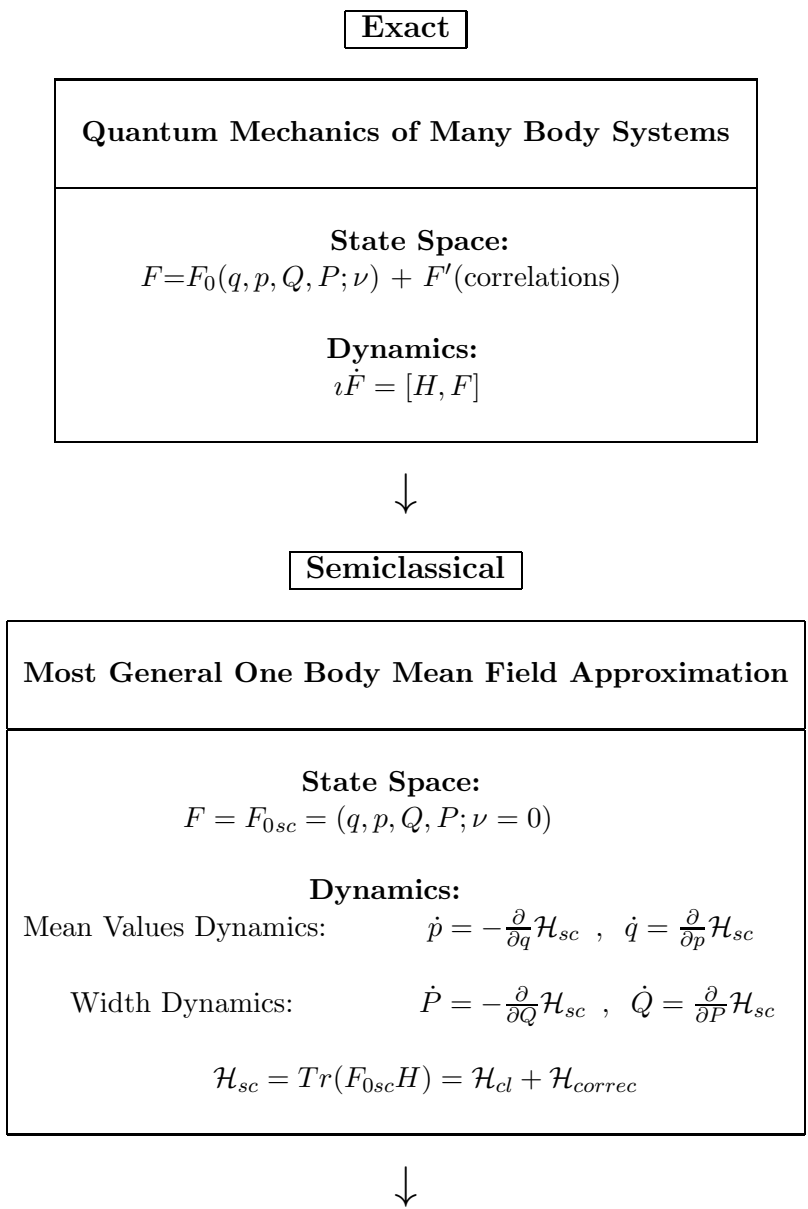
\begin{center}
\begin{tabular}{|l|}
\hline
{\large\bf{Exact}}\\
\hline
\end{tabular}
\end{center}

\begin{center}
\begin{tabular}{|l|}
\hline
\\
\bf{ Quantum Mechanics of Many Body Systems } \\
\\
\hline
\\
 
\hspace{3.3cm}\bf{ State Space:} \\

\hspace{0.8cm} 
$F$=$F_0(q,p,Q,P;\nu)$ + $F'$(correlations)\\
\\

\hspace{3.4cm}\bf{Dynamics:} \\

\hspace{3.5cm}$ \imath \dot{F} =[H,F] $ \\
\\
\hline
\end{tabular}

\end{center}

\begin{center}
{\LARGE{\bf $\downarrow$}}
\end{center}

\begin{center}
\begin{tabular}{|l|}
\hline 
{\large\bf{Semiclassical}}\\
\hline
\end{tabular}
\end{center}

\begin{center}
\begin{tabular}{|l|}
\hline
\\
\bf{Most General One Body Mean Field Approximation} \\
\\
\hline
\\
 
\hspace{3.7cm}\bf{State Space:} \\

\hspace{2.0cm} $F= F_0$$ _{sc}=(q,p,Q,P;\nu=0)$\\
\\

\hspace{3.8cm}\bf{Dynamics:} \\

Mean Values Dynamics: \hspace{1.0cm}
$\dot{p}=-\frac{\partial}{\partial q} {\cal{H}}_{sc} \;\; , \;\;
\dot{q}=\frac{\partial}{\partial p} {\cal{H}}_{sc} $\\

\\
\hspace{0.5cm} Width Dynamics:\hspace{1.5cm}
$\dot{P}=-\frac{\partial}{\partial Q} {\cal{H}}_{sc} \;\; , \;\;
\dot{Q}=\frac{\partial}{\partial P} {\cal{H}}_{sc}$ \\
\\
\hspace{2.4cm}
${\cal{H}}_{sc}=Tr(F_{0sc} H)={\cal{H}}_{cl} + {\cal{H}}_{correc}$\\
\\
\hline
\end{tabular}
\end{center}

\begin{center}
{\LARGE{\bf $\downarrow$}}
\end{center}

\begin{center}
\begin{tabular}{|l|}
\hline
{\large\bf{Classical}}\\
\hline
\end{tabular}
\end{center}

\begin{center}
\begin{tabular}{|l|}
\hline
\\
\bf{One Body Coherent State Approximation } \\
\\
\hline
\\
 
\hspace{3.3cm}\bf{Kinematics:} \\

\hspace{2.5cm} 
$F=F_{0cl}(q,p;\nu=0)$\\
\\

\hspace{3.4cm}\bf{Dynamics:} \\

\hspace{2.4cm}
$\dot{p}=-\frac{\partial}{\partial q} {\cal{H}}_{cl} \;\; , \;\;
\dot{q}=\frac{\partial}{\partial p} {\cal{H}}_{cl} $\\
\\
\hspace{2.9cm}
${\cal{H}}_{cl}=Tr(F_{0cl} H)$\\
\\
\hline
\end{tabular}

\end{center}

\vspace{0.5cm}

\noindent
where $F_{0cl}$ is obtained from eq.(13) and the Bogoliubov transform (4)
by setting $x=1$, $y=0$ and $\nu=0$, {\it i.e.},  $F_{0cl}$ corresponds
to a coherent minimum uncertainty state. On the other hand  $F_{0sc}$ 
incorporates the dynamics of the quadratures $<a^{\dag}a>$ and 
$<a^{\dag}a^{\dag}> $ related to the variances (eqs. (11) and (12)) 
and is a pure
state ($\nu=0$). The dynamics of $<a^{\dag}a>$ and $<a^{\dag}a^{\dag}>$
enter as a correction $({\cal{H}}_{correc})$ to the classical one
$({\cal{H}}_{cl})$.
 Of course considering $\nu \neq 0$ would enlarge the class
 of gaussian states so as to encompass mixtures. Since there are no rigorous
 classical results available which cover this generality we restrict 
ourselves to $\nu=0$.

Since on the one body mean field approximation
the occupation probability $\nu$ does not depend on time  we have
an even number of parameters which characterize the time evolution of $F_0$, namely
$q,p,x$ and $y$ - in this case $\nu$ enter the dynamics as a free parameter. 
 This enable one to  cast (in a simple way) the variances  $\Delta q$ and
$\Delta p$ in the form of canonical variables $(Q,P)$ \cite{squeezing} quoted
in the above scheme. We define

\begin{equation}
x=\cosh{\sigma} + \imath \frac{\tau}{2} \;\;\;,
\end{equation}

\begin{equation}
y = \sinh{\sigma} + \imath \frac{\tau}{2}  
\;\;\;,
\end{equation}

and then 

\begin{equation}
P=\sqrt{ \frac{(1+2 \nu)}{2}} \;\; \tau \;\;\; ,
\end{equation}

\begin{equation}
Q=\sqrt{ \frac{(1+2 \nu)}{2}} \;\; e^{- \sigma} \;\;\; .
\end{equation}

In the semiclassical level, defined by $F_{0sc}$, the transformations are
 the same,
setting $\nu=0$. The connection of these semiclassical variables with the 
variables used in refs. \cite{Pattanyak2} and \cite{Jackiw} is made by a
 simple  transformation

\begin{equation}
G=Q^2\;\;\;\;\; ; \;\;\;\;\; \Pi=\frac{P}{2Q} \;\;\;\;\;\;\;\;; \nu=0 \;\;\; .
\end{equation}

It is a simple matter to check that eqs. (11) and (12),
in terms of these variables  acquire a very simple interpretable
form 

\begin{equation}
\Delta q^2 = Q^2 \;\;\; ,
\end{equation}

\begin{equation}
\Delta p^2 = P^2  + \frac{(1+2 \nu)^2}{4 Q^2} \;\;\; .
\end{equation}

It becomes clear how the uncertainty 
principle will manifest in the semiclassical approximation:
The fact that the width of the wave packet cannot be zero is
translated in classical terms by a centrifugal barrier

\begin{equation}
\frac{(1+2 \nu)^2}{4 Q^2} \;\;\; .
\end{equation}

\noindent
of course
this term  will be introduced also into the dynamics and coupled to the other 
degrees of freedom.  The uncertainty product  is given by

\begin{equation}
\Delta q\Delta p=
\sqrt{\frac{(1+2 \nu)^2}{4}+P^2Q^2} \;\; \geq\ \;\; \frac{(1+2 \nu)}{2} 
\;\;\; ,
\end{equation}

\noindent
and  the minimal uncertainty  situation in the limit $\nu \rightarrow 0$ 
is  $\Delta q=\Delta p=\sqrt {\frac{1}{2}}$.
We get it  with $P=0$ and $Q=\sqrt {\frac{1}{2}}$.

The presence of the centrifugal barrier in equation (25)  shows
that it came from the gaussian approximation in a purely
 kinematical way and therefore does not depend on the dynamics $H$.

The next step  is to obtain the time evolution of the complete set of
gaussian parameters. This is done by means of the quantum equation 
of motion

\begin{equation}
\imath \frac{d}{dt} <{\cal{O}}  >=
Tr (F [{\cal{O}},H]) \;\;\; ,
\end{equation}

\noindent
where ${\cal{O}}$ is any operator we have chosen as relevant.
Let us  calculate the  l.h.s. of eq. (26).

We calculate the time derivative of the matrix $\bf{N}$ (eq. (8))
\noindent

\begin{equation}
\bf{N}= \bf{X^{\dag} R X}=
\left(
\begin{array}{cc}
\nu & 0 \\
0 & 1 + \nu 
\end{array}
\right) \;\;\;,
\end{equation}

\begin{equation}
\frac{d}{dt} {\bf{N}} =\frac{d}{dt} {\bf {X^{\dag} R X}}  
\end{equation}

\noindent
which gives

\begin{equation}
\bf{ X^{\dag} \dot{R} X = \dot{N} - 
\dot{X}^{\dag} R X - X^{\dag}R \dot{X}} \;\;\;,
\end{equation}

\noindent
and with help of the normalization condition $\bf{ X^{\dag}G X}=G$,

\begin{equation}
\bf{ X^{\dag}\dot{R}X=\dot{N}-\dot{X}^{\dag}GXGN-NGX^{\dag}G\dot{X}}
\;\;\;.
\end{equation}

Writing explicitly the above matrix equation and comparing them element by 
element, we obtain 

\begin{equation}
 i \frac{d}{dt}  \tilde{\eta}^{\dag} \tilde{\eta}^{\dag} =
 i (\dot{x} y -x \dot{y})( 1 + 2 \nu) = 
Tr([\tilde{\eta}^{\dag} \tilde{\eta}^{\dag},H]F_0) \;\;\; .
\end{equation}

For the condensate $<a >$ the equation of motion is obtained
directly from the parameterization (3) and  the Bogoliubov transformation  as

\begin{equation}
i\frac{d}{dt} \{  \sqrt{ \frac{\mu_0}{2} } ( q+ \frac{i}{\mu_0} p) \}
= x Tr([\eta,H]F_0) - y^* Tr([\eta^{\dag},H] F_0) \;\;\; .
\end{equation} 

The Hamiltonian dynamics in terms of canonical variables is obtained
with transformations (17)-(20) which splits eqs. (30) and (31) into
a set of four equations for the corresponding real quantities $q,p,Q$ 
and $P$. 
  
\newpage

\section{The Model and the Semiclassical Dynamics:
 1/N corrections}

The $SU(2)$ Lipkin Model \cite{Lipkin} 

\begin{equation}
H= \epsilon J_z + \frac{V}{2}(J_+^2 + J_-^2) \;\;\; ,
\end{equation}

\noindent
is characterized in Schwinger's 
representation by \cite{RingSchuck}

\begin{equation}
H = \frac{1}{2} \epsilon (b^{\dag}b - a^{\dag}a ) + \frac{V}{2}
[b^{\dag} b^{\dag} a a + a^{\dag} a^{\dag} bb ] \;\;\; ,
\end{equation}

\noindent
where $b^{\dag}$ and $a^{\dag}$ are the creation operators for bosons 

\begin{equation} 
[a,a^{\dag}]=1 \hspace{4.0cm} [b,b^{\dag}]=1 \;\;\; .
\end{equation}

\noindent
Here we have $\hbar=1$. 
The realization of SU(2) algebra in this representation is

\begin{equation}
J_z= \frac{1}{2} (b^{\dag}b- a^{\dag}a) \;\;\; ,
\end{equation}

\begin{equation}
J_+= b^{\dag}a \;\;\;\;\;\;\; , \;\;\;\;\;\;\; J_-=  a^{\dag} b.
\end{equation}

The number of particles in the system is given by 
$N= b^{\dag}b+ a^{\dag}a $ and the Casimir operator $J^2$ have eigenvalues 
equal to $N/2(N/2 +1)$, with $N=2J$. 

In order to obtain the mean field approximation we first define
 the normalized density matrix 
$F_0(\tilde\alpha^{\dag} \tilde \alpha,\tilde\beta^{\dag} \tilde \beta; 
\nu_\alpha,\nu_\beta) $, {\it i.e.}, encompassing mixtures. The corresponding
semiclassical approximation may be immediately obtained by setting $\nu_i=0$.
We have

\begin{equation}
F_0 =F_{0 \alpha} \otimes F_{0 \beta} \;\;\; ,
\end{equation}

\noindent 
where 

\begin{equation}
F_{0 \alpha}=\frac{1}{1+\nu_\alpha} 
(\frac{\nu_\alpha}{1+\nu_\alpha})^{\tilde{\alpha}^{\dag} \tilde{\alpha}}
\;\;\;\;\; ; \;\;\;\;\;
F_{0 \beta}=\frac{1}{1+\nu_\beta}
(\frac{\nu_\beta}{1+\nu_\beta})^{\tilde{\beta}^{\dag} \tilde{\beta}}
\end{equation}

\noindent and

\begin{equation}
\tilde\alpha=x_a (a - <a >)+
 y_a (a^{\dag} - < a^{\dag} >) \;\;\; ,
\end{equation}
 
\begin{equation}
\tilde\beta=x_b(b-< b>)+y_b(b-< b^{\dag}>)
\;\;\; .
\end{equation}

The next tedious but straightforward step is to rewrite the 
Hamiltonian (33) in terms of the operators in the equations above
$(H(a^{\dag}a,a^{\dag}a^{\dag},aa,b^{\dag}b,b^{\dag}b^{\dag},bb)$
$\rightarrow$ 
$H(\tilde\alpha^{\dag}\tilde\alpha,\tilde\alpha^{\dag} \tilde\alpha^{\dag},
\tilde\alpha \tilde\alpha,\tilde{\beta}^{\dag} \tilde{\beta},
\tilde{\beta}^{\dag} \tilde{\beta}^{\dag},\tilde{\beta} \tilde{\beta} ))$ 
so that the necessary traces are easily calculated. We get

\begin{equation}
\imath\dot{\nu}_{\alpha}=
\imath \frac{d}{dt}<\tilde{\alpha}^{\dag}\tilde{\alpha}>=
Tr{F_0 [\tilde{\alpha}^{\dag} \tilde{\alpha},H]}=0
\end{equation}

\begin{equation}
\imath \dot{\nu}_\beta=
\imath \frac{d}{dt}<\tilde{\beta}^{\dag} \tilde{\beta}>=
Tr{F_0 [\tilde{\beta}^{\dag} \tilde{\beta},H]}=0
\end{equation}

\[
i \frac{d}{dt}< \tilde{\alpha}^{\dag} \tilde{\alpha}^{\dag}>=
Tr{F_0 [\tilde{\alpha}^{\dag} \tilde{\alpha}^{\dag},H]} \Rightarrow
\]
\begin{equation}
\imath(\dot{x}_a y_a - x_a \dot{y}_a)(1+2 \nu_\alpha)=
-\epsilon x_a y_a (1+ 2\nu_\alpha)  +
\end{equation}
\[
- V x_a^2 (1+ 2\nu_\alpha) [x^{*2}_b   <\beta^{\dag}> ^2 + y^{2}_b <\beta>^2
-x^*_b y_b( 2 \nu_\beta+  2  <\beta^{\dag}> <\beta> +1)] +
\]
\[
- V y_a^2 (1+ 2\nu_\alpha) [ x^{2}_b <\beta>^2 + y^{*2}_b <\beta^{\dag}> ^2 
-x_b y^{*}_b (2 \nu_\beta+ 2  <\beta^{\dag}> <\beta>+1)] \;\;\; ,
\]

\[
i \frac{d}{dt}<\tilde{\beta}^{\dag} \tilde{\beta}^{\dag}>=
Tr{F_0 [\tilde{\beta}^{\dag} \tilde{\beta}^{\dag},H]}  \Rightarrow
\]
\begin{equation}
\imath (\dot{x}_b y_b - x_b \dot{y}_b)(1+2 \nu_\beta)=
-\epsilon x_b y_b (1+ 2\nu_\beta) +
\end{equation}
\[
- V x_b^2 (1+ 2\nu_\beta) [x^{*2}_a   <\alpha^{\dag}> ^2 + y^{2}_a <\alpha>^2
-x^*_a y_a(   2\nu_\alpha+  2  <\alpha^{\dag}> <\alpha> +1)] +
\]
\[
- V y_b^2 (1+ 2\nu_\beta) [ x^{2}_a <\alpha>^2 + y^{*2}_a <\alpha^{\dag}> ^2 
-x_a y^{*}_a (  2\nu_\alpha+ 2  <\alpha^{\dag}> <\alpha>+1)]  \;\;\; .
\]

From equations (42) and (43) we see that $\dot \nu_a= \dot \nu_b=0$. This is
directly a consequence of the mean field approximation. Thus in the
semiclassical level, $\nu_i$ enters the dynamics as a free parameter.
For the bosonic condensate the equations of motion read
\begin{equation}
i\frac{d}{dt}<a> = x_a Tr[\alpha,H]F - y^*_a tr[\alpha^{\dag},H] F_0 \;\;\; ,
\end{equation} 

\begin{equation}
i\frac{d}{dt}<b> = x_b Tr[\beta,H]F - y^*_b tr[\beta^{\dag},H] F_0 \;\;\; ,
\end{equation} 

\noindent
where
\begin{equation}
Tr{ F_0 [\alpha^{\dag},H] }= - \frac{1}{2} \epsilon
[2 x_a y_a <\alpha>  -( x^*_a x_a + y^*_a y_a)<\alpha^{\dag}>] +
\end{equation}
\[ 
-V(y_a^2 <\alpha> - y_a x^*_a<\alpha^{\dag}>)*
\]
\[
*[ x^{2}_b <\beta>^2 + y^{*2}_b <\beta^{\dag}> ^2 
-x_b y^{*}_b ( 2 \nu_\beta + 2  <\beta^{\dag}> <\beta>+1]
\]
\[
-V(x_a^2 <\alpha> - x_a y^*_a<\alpha^{\dag}>)*
\]
\[
*[x^{*2}_b   <\beta^{\dag}> ^2 + y^{2}_b <\beta>^2
-x^*_b y_b( 2 \nu_\beta  + 2  <\beta^{\dag}> <\beta> +1)]
\]

\begin{equation}
Tr{F_0 [\beta^{\dag},H]}=- \frac{1}{2} \epsilon
[ 2 x_b y_b <\beta> - ( x_b^*x_b + y_b^* y_b) <\alpha^{\dag}>]+
\end{equation}
\[
-V(x_b^2 <\beta> - x_b y^*_b <\beta^{\dag}>)*
\]
\[
*[x^{*2}_a   <\alpha^{\dag}> ^2 + y^{2}_a <\alpha>^2
-x^*_a y_a( 2 \nu_\alpha  + 2  <\alpha^{\dag}> <\alpha> +1)]
\]
\[
-V( y_b^2<\beta> - x^*_b y_b <\beta^{\dag}>)*
\]
\[
*[ x^{2}_a <\alpha>^2 + y^{*2}_a <\alpha^{\dag}> ^2 
-x_a y^{*}_a ( 2 \nu_\alpha + 2  <\alpha^{\dag}> <\alpha>+1)]
\]

Now the formal definition of the classical limit for $1/N$  type models is in
order.

From the $SU(2)$ quasispin operators which characterize the algebra of the
 model one may write \cite{Galleti}

\begin{equation}
[{\cal{J}}_i,{\cal{J}}_j]=\frac{i \hbar}{J} \epsilon_{ i j k}{\cal{J}}_k \;\; 
\end{equation}

\noindent
where ${\cal{J}}_i= J_i/J$ and  $J= N/2$. The classical limit is 
mathematically defined as

\begin{equation}
\lim_{J \rightarrow \infty} \frac{J}{i \hbar}
[{\cal{J}}_i,{\cal{J}}_j]=  \epsilon_{ i j k}{\cal{J}}_k \;\;\; .
\end{equation}

For finite $J$ the eigenvalues of the ${\cal{J}}_i '$s are mapped into the 
interval $[-1,1]$ and the spectra get denser as $J$ increases while $\hbar /J$
decreases. In Schwinger's representation this procedure implies that the 
bosonic operators $a$ and $b$ be scaled as \cite{Van Hemmen}

\begin{equation}
a^{(\dag)} \rightarrow \frac{ a^{(\dag)} }{\sqrt{J}}
\end{equation}

\begin{equation}
b^{(\dag)} \rightarrow \frac{ b^{(\dag)} }{\sqrt{J}}\;\;\;.
\end{equation}

Since we have the Hamiltonian as a function of the Bogoliubov operators 
$H(\tilde\alpha^{\dag}\tilde\alpha,\tilde\alpha^{\dag} \tilde\alpha^{\dag},
\tilde\alpha \tilde\alpha,\tilde{\beta}^{\dag} \tilde{\beta},
\tilde{\beta}^{\dag} \tilde{\beta}^{\dag},\tilde{\beta} \tilde{\beta} ))$ 
writing down the corresponding equations of motion in the canonical variable 
phase space is straightforward. Firstly we invert the Bogoliubov
 transformation to obtain from parameterization (3) the mean values
$<\alpha> $ and $<\beta>$ as a function of the scaled quantities
$q_i'= q_i/\sqrt{J}$ and $p_i'=p_i/\sqrt{J}$ $(i=a,b)$.

\begin{equation}
<\alpha^{\dag}>=\sqrt{\frac{1}{2}}[q_a' (x_a +y_a)-ip_a' (x_a - y_a)] \;\;\;,
\end{equation}

\begin{equation}
<\beta^{\dag}>=\sqrt{ \frac{1}{2}} [ q_b' (x_b +y_b) - i p_b' (x_b - y_b)]
\hspace{0.2cm} \;\;\; .
\end{equation}
\noindent

Then we use equations (17)-(20) writing the Bogoliubov
 parameters $x_i$ and $y_i$
in terms of $Q_i, P_i$ and $\nu_i$. In this way we get

\[
{\cal{H}}_{sc}(q_i',p_i',Q_i,P_i)= {\cal{H}}_{cl} (q_i',p_i') +
\]
\begin{equation}
+ \frac{1}{J} {\cal{H}}_{correc}^1 (q_i',p_i',Q_i,P_i;\nu_i=0) + 
 \frac{1}{J^2} {\cal{H}}_{correc}^2 (Q_i,P_i;\nu_i=0) \;\;\;,
\end{equation}
\noindent
where
\begin{equation}
{\cal{H}}_{cl}(q_i',p_i')=\frac{1}{2} \epsilon 
 \{ \frac{1}{2}(q_b'^2 + p_b'^2) - \frac{1}{2}(q_a'^2 + p_a'^2) \} + 
\end{equation}
\[
\chi \{ \frac{1}{2}(q_a'^2 - p_a'^2) \frac{1}{2}(q_b'^2 - p_b'^2)
+  q_b' p_b' q_a' p_a' \} \;\;\; ,
\]

\begin{equation}
{\cal{H}}_{correc}^1 (q_i',p_i',Q_i,P_i)=\frac{1}{2} \epsilon 
\{ \frac{1}{2} (Q_b^2+ P_b^2) - \frac{1}{2} ( Q_a^2 + P_a^2) +
\frac{\xb}{8 Q_b^2} - \frac{\xa}{8 Q_a^2} \} +
\end{equation}
\[
-\chi \{ 
 \frac{1}{2}(q_b'^2-p_b'^2)[\frac{\xa}{8Q_a^2}-
\frac{1}{2}(Q_a^2-P_a^2)]+q_b'p_b'Q_aP_a \} +
\]
\[
-\chi \{ 
\frac{1}{2}(q_a'^2-p_a'^2)[\frac{\xb}{8 Q_b^2}-
\frac{1}{2}(Q_b^2-P_b^2)]+q_a'p_a'Q_bP_b \} \;\;\; ,
\]

\begin{equation}
 {\cal{H}}_{correc}^2 (Q_i,P_i)= \chi \{  
 [ \frac{\xb}{8 Q_b^2} -\frac{1}{2}(Q_b^2-P_b^2) ]
 [ \frac{\xa}{8 Q_a^2} -\frac{1}{2}(Q_a^2-P_a^2) ] 
+   Q_b P_b Q_a P_a \}\;\;\; .
\end{equation}

\noindent
In the above equations $\chi= V (2 J)$ is the scaled interaction parameter.

In the  limit $J \rightarrow \infty$  the scaled number of particles 

\[
{\cal{N}} =Tr [(a^{\dag}a +b^{\dag}b  )F]/J \;\; ,
\]
\begin{equation}
{\cal{N}}= \frac{1}{2}(q_b'^2+p_b'^2) + \frac{1}{2}(q_a'^2+p_a'^2) +
\frac{1}{J} 
[ \frac{\xb}{8 Q_b^2} +\frac{1}{2}(Q_b^2+P_b^2)] +
\frac{1}{J}
[ \frac{\xa}{8 Q_a^2} +\frac{1}{2}(Q_a^2+P_a^2)]  \;\; ,
\end{equation}

\noindent
is a constant of  motion. We have
\begin{equation}
\frac{d}{dt}{\cal{N}}= \{ {\cal{H}} ,{\cal{N}} \} =0 \;\; ,
\end{equation}

\noindent
with the Poisson brackets defined as
\begin{equation}
\{ f(q_i,p_i,Q_i,P_i), g(q_i,p_i,Q_i,P_i) \} = \sum_i 
\frac{\partial f}{\partial q_i} \frac{\partial g}{\partial p_i} -
 \frac{\partial g}{\partial q_i} \frac{\partial f}{\partial p_i} +     
\frac{\partial f}{\partial Q_i} \frac{\partial g}{\partial P_i} -
 \frac{\partial g}{\partial Q_i} \frac{\partial f}{\partial P_i} \;\; .
\end{equation}

\noindent
Although if $J$ is finite, the Poisson bracket 
$\{ {\cal{H}}_{sc} ,{\cal{N}} \} $ do not vanishes and we have 
$\frac{d}{dt} {\cal{N}}  \neq 0$.

As is well known the number of particles $N=2 J$ plays the role of a 
semiclassical expansion parameter in the $SU(2)$ Lipkin Model \cite{lipclass}.
Therefore in the limit $N \rightarrow \infty$ the
Hamiltonian   ${\cal{H}}_{cl}(q_i,p_i)$  precisely 
corresponds to the classical limit of the model usually taken by
means of coherent states \cite{lipclass}.

The term ${\cal{H}}_{correc}^1 (q_i',p_i',Q_i,P_i)$ in eq.(56) 
which is the first 
order quantum correction  contains 
the dynamics of the quadratures coupled  to the mean values. Notice that
for the minimal uncertainty initial condition 
$\{q_i'(0),p_i'(0), Q_i(0) =\sqrt{\frac{1}{2}} , P_i(0)=0\}$,
we have the total energy equal to the classical one 
${\cal{H}}_{sc}={\cal{H}}_{cl}|_{t=0}~$ $({\cal{H}}_{correc}^1|_{t=0}=0)$.
 The term ${\cal{H}}_{correc}^2 (Q_i,P_i)$ contains  only the dynamics of the
 quadratures, and if one chose an initial condition with the minimal
 uncertainty situation   it does not 
contribute to the  dynamics  ${\cal{H}}_{correc}^2(t)=0$ 
(for the sake of clarity we now reset the variables 
$ q_i', p_i' \rightarrow q_i,p_i$).

The  equations of motion on the $\{q_i,p_i,Q_i,P_i\}$ phase space  
are obtained as follows: we
use (17)-(20) to rewrite the l.h.s. of eqs. (44) and (45) in terms of 
$\dot{Q}_i$ and $\dot{P}_i$  and parameterization (3) to rewrite the 
l.h.s. of eqs. (46) and (47) in terms of 
$\dot{q}_i$ and $\dot{p}_i$. Doing the same transformations on the r.h.s.
of (44)-(47) 
 and comparing their real and imaginary parts the two equations for each
level $a$ and $b$ split into a set of four real equations for the 
corresponding real quantities $\dot{q}_i,\dot{p}_i ,\dot{Q}_i$ and $\dot{P}_i$
($i=a,b$). 

\begin{equation}
\dot{q}_a= -\frac{\epsilon}{2}p_a +
 \chi [ q_a ( \frac{Q_b P_b}{J} + q_b p_b) - p_a \frac{1}{2}(q_b^2 - p_b^2)]+
\end{equation}
\[
 +\chi p_a [ \frac{\xb}{8JQ_b^2} - \frac{1}{2J}(Q_b^2 -P_b^2)] \;\;,
\]

\begin{equation}
\dot{p}_a=\frac{\epsilon}{2}q_a -
 \chi [p_a( \frac{Q_b P_b}{J} + q_b p_b) + q_a \frac{1}{2} (q_b^2 - p_b^2)] +
\end{equation}
\[
+ \chi q_a [ \frac{\xb}{8JQ_b^2} - \frac{1}{2J}(Q_b^2 -P_b^2)]  \;\;,
\]

\begin{equation}
\dot{q}_b= \frac{\epsilon}{2}p_b +
 \chi [ q_b (  \frac{Q_a P_a}{J}+ q_a p_a) - p_b \frac{1}{2}(q_a^2 - p_a^2)]+
\end{equation}
\[
 +\chi p_b [  \frac{\xa}{8JQ_a^2} - \frac{1}{2J}(Q_a^2 -P_a^2)] \;\;,
\]

\begin{equation}
\dot{p}_b=- \frac{\epsilon}{2}q_b -
 \chi [p_b( \frac{Q_a P_a}{J} + q_a p_a) + q_b \frac{1}{2} (q_a^2 - p_a^2)] +
\end{equation}
\[
+\chi q_b [ \frac{\xa}{8JQ_a^2} - \frac{1}{2J}(Q_a^2 -P_a^2)] \;\;,
\]   

\begin{equation}
\dot{Q}_a=-\frac{\epsilon}{2J}P_a+\chi \frac{P_a}{J^2}[\frac{\xb}{8Q_b^2}
 -\frac{1}{2}(Q_b^2 -P_b^2)] +
\end{equation}
\[
+\chi[ \frac{Q_a}{J}(\frac{ Q_b P_b}{J} + q_b p_b) -
 \frac{P_a}{J} \frac{1}{2}(q_b^2 - p_b^2)]  \;\;,
\]

\begin{equation}
\dot{P}_a= \frac{\epsilon}{2J} ( Q_a - \frac{\xa}{4Q_a^3}) +
\end{equation}
\[
+\chi \{\frac{1}{J^2} [ \frac{\xa}{4Q_a^3} + Q_a]
 [ \frac{\xb}{8Q_b^2} -\frac{1}{2}(Q_b^2 -P_b^2)]+
\]
\[
- \frac{1}{2J}(q_b^2 - p_b^2)[\frac{\xa}{4Q_a^3} + Q_a]
-\frac{P_a}{J} ( \frac{Q_b P_b}{J} + q_b p_b) \} \;\;,
\]

\begin{equation}
\dot{Q}_b = + \frac{\epsilon}{2J}P_b +\chi\frac{P_b}{J^2}[\frac{\xa}{8Q_a^2} 
-\frac{1}{2}(Q_a^2 -P_a^2)]
 +
\end{equation}
\[
+  \chi [\frac{ Q_b}{J}(  \frac{Q_a P_a}{J} + q_a p_a) -
 \frac{P_b}{J} \frac{1}{2}(q_a^2 - p_a^2)]  \;\;,
\]

\begin{equation}
\dot{P}_b= \frac{\epsilon}{2J} ( \frac{\xb}{4Q_b^3}-Q_b) +
\end{equation}
\[
+\chi \{ \frac{1}{J^2} [ \frac{\xb}{4Q_b^3} + Q_b]
 [ \frac{\xa}{8Q_a^2} -\frac{1}{2}(Q_a^2 -P_a^2)]+
\]
\[
- \frac{1}{2J}(q_a^2 - p_a^2)[\frac{\xb}{4Q_b^3} + Q_b]
-\frac{P_b}{J} (\frac{Q_a P_a}{J} + q_a p_a) \} \;\;\; .
\]

We next analyze the consequences of these $1/N$ corrections. 

\newpage

\section{Results:}

\vspace{1.0cm}

What are then the effects of the
dynamics of the width  (and therefore the uncertainty principle) on the 
classical dynamics of the Lipkin Model? We will discus our results 
 both qualitatively and quantitatively.

Firstly the semiclassical manifestation of the uncertainty principle is the 
appearance of  new degrees  of freedom whose position $Q_i$ cannot be zero. 
This is formally achieved by a repulsive centrifugal-type potential. 
  The inclusion of these new degrees of freedom destroys the
integrability of the classical dynamics and  chaotic behavior emerges. 
Since the Lipkin Model  has a regular motion both in the quantum regime 
and classical limit the semiclassical chaos arises as an artifact 
(a legitimate one) of the 
approximation. The main result we want to stress here is that despite of 
inducing chaotic behavior quantum  ($1/N$) correction gives a better 
description of the  time evolution of observables. Other interesting
feature is: Chaos is the mechanism through which quantum properties are 
effected on the semiclassical phase space (such as quantum tunneling 
effect).  
 
Let us now describe the classical dynamics in the four dimensional
phase space. Beside the conservation of
energy ${\cal{H}}_{cl}$ we have also the constraint in ${\cal{N}}$. Thus,
once it is fixed, there exists one, and only one trajectory 
satisfying ${\cal{N}}=<N>/(J)=2$ 
with a given value of ${\cal{H}}_{cl}$. The existence of these two constants 
of motion enables one  to show in the same
Poincar\'e section the trajectories for all available energies for a given
value of the interaction parameter $\chi$. In figure 1 we show the well 
known second order phase transition exhibited by the model in its 
classical limit.
For any value of $\chi$ below the critical one $|\chi| <|\chi_{crit}|=1.0$,  
the invariant tori are all of the same kind and represent the 
{\it rotational} aspect of the dynamics (see figure 1(a)). In this case 
the possible range of energies is $|{\cal{H}}_{cl}|<1.0$. In figure 1(a) 
the energies
 increase from the  boundary (where ${\cal{H}}_{cl}=-1.0$) to the origin
 (where ${\cal{H}}_{cl}=1.0$).
For $|\chi| >|\chi_{crit}|=1.0$ the possible range of energies is enlarged
$|{\cal{H}}_{cl}|<|(1+ \chi^2)/(2 \chi)|$.
We still have the {\it rotating} trajectories $|{\cal{H}}_{cl}|<1.0$   
(which we label by E$_{rot}$ in figure 1(b)) and  we also have the 
{\it deformed}
ones $-1.0<{\cal{H}}_{cl}<(1+ \chi^2)/(2 \chi)$ and 
 $1.0<{\cal{H}}_{cl}<-(1+ \chi^2)/(2 \chi)$ which we label by E$_{min}$
and E$_{max}$ respectively in figure 1(b) ($\chi=-6.0$).
 The fixed points of the 
Poincar\'e map associated to the extreme energies are 

\begin{eqnarray} 
q_a &= & 0\;\;\;\;\;\;;\;\;\;\;\;\; p_a
= \pm \sqrt{( 1- \frac{1}{\chi})}\nonumber\\ 
q_b & = & 0\;\;\;\;\;\;;\;\;\;\;\;\; p_b=\pm  \sqrt{( 1+ \frac{1}{\chi})}\;,
\end{eqnarray}

\noindent
which give us 
\[
{\cal{H}}_{cl}=\frac{1}{2}\frac{1+ \chi^2}{\chi}\;\;\;,
\]

\noindent 
and

\begin{eqnarray} 
q_a  &=& \pm \sqrt{(1+\frac{1}{\chi})} \hspace{0.5cm};\hspace{0.5cm}p_a=0
\nonumber\\
q_b  &=& 0 \hspace{2.3cm} ; \hspace{.5cm} p_b= \pm\sqrt{( 1- \frac{1}{\chi})} 
\end{eqnarray}

\noindent
which gives 
\[
{\cal{H}}_{cl}=-\frac{1}{2}\frac{1+ \chi^2}{\chi}\;\;\;.
\]

The {\it rotational} trajectories are isolated from the {\it deformed} ones
by two separatrices  S$_-~$ (${\cal{H}}_{cl}$(S$_-$)=-1.0)
and S$_+~$ (${\cal{H}}_{cl}$(S$_+$)=1.0). See figure 1(b).

Introducing  the correction terms in ${\cal{H}}_{sc}$ , and 
therefore the new degrees 
of freedom related to widths $(Q_i,P_i)$ coupled to mean values $(q_i,p_i)$, 
the geometrical  structure of  the integrable system is 
destroyed and 
chaotic behavior  emerges. The quantity ${\cal{N}}$ is not a constant 
of motion any more. For increasing $J$ the integrability of the 
classical limit  is gradually recovered (see figure 2(a)-(c)) and takes 
place again  only in the limit $J \rightarrow \infty$.
  
Of course in the classical domain tunneling effects are completely forbidden.
However this it is not
the case when the quantum corrections we are dealing with are taken into 
account. Let us now look at  the low energy orbits 
$(1+\chi^2)/(2 \chi)<{\cal{H}}_{cl}<-1.0$ ($\chi=-6.0$) .
 The classically corresponding invariant 
tori are localized in the two symmetric regions  E$_{min}$ with
$p_a<0$ or $p_a>0$ exclusively and because of classical integrability 
the time evolution  of any chosen initial condition  will be  confided on 
its respective region. Therefore, destroying integrability is
the way the quantum correction works to effect quantum tunneling.
Choosing an initial condition with energy ${\cal{H}}_{cl}<-1.0$ 
in the  $p_a < 0$ semiplane as an example we show in figure 3(a) its 
classical Poincar\'e section. For  finite values of $J$ this same 
initial condition (evolved semiclassically) is able to access the 
symmetrical region $p_a >0$ 
(figure 3(b)). The quantum observable associated to the transition 
between the semiplanes $p_a<0$ and $p_a>0$ is $J_x=(J_+ +J_-)/2$. 
Its mean value sign on the Poincar\'e section ($q_b=0$ , $p_b>0$) 
is determined by the $p_a$ sign.

\[
 Sign [<J_x>]=Sign[\frac{1}{2}(q_b q_a + p_b p_a)]= Sign[p_a] \;\;\;.
\]

The frequency of such transitions also depends on energy, increasing
for energies near the separatrix and decreasing as $J$ increase.
A quantitative measure of this process may be achieved by studying its 
time scale. We define the confinement time  $T_c$ as  the time interval 
between two transitions   $p_a<0 \leftrightarrow p_a>0~~~$.
 We then  divide
it by the Poincar\'e time $T_p$ ,{\it i.e.}, the amount of time required 
by starting with an initial condition on the Poincar\'e plane and evolving 
 until it
reaches the plane again.  We  take the average  over the first thousand
values of  $\bar{ T }_c$ and evaluate the dimensionless quantity 
$\bar{ T }_c/ / \bar{ T }_p$ as a function of $J$ for different energies
(see figure 4). 
This  quantity can also be interpreted as the average number
of iterations of the Poincar\'e map necessary for  a transition to occur. 
From the figure we note that the transitions become scarce as $J$ increase.
For any finite value of $J$ there must occur a transition, although, for large
values of $J$ (or energies close to the minimum value) this may require 
numerically integrating the equations of motion for
an {\it enormous } amount of time.
 Another interesting feature in figure 4 is the dependence
on energy. For an  energy close to the separatrix 
(${\cal{H}}_{cl}=-1.01$) $\bar{ T }_c / \bar{ T }_p$ increases slower than
for lower energies (${\cal{H}}_{cl}=-1.1$ and ${\cal{H}}_{cl}=-1.2$).
Since  tunneling effect manifests itself through  chaotic motion it must be
more conspicuous where chaos (roughly speaking) persists, {\it i.e.},
 near the separatrix.

Notice that here we do not intend  to rigorously define a tunneling rate 
(in terms of energy splittings), which is an 
interesting problem in itself in particular for spin systems, we are just
characterizing the phenomenon in the gaussian representation. 

An important question that naturally arises in this work concerns the time
evolution of observables. Since the nonintegrability of the semiclassical
description is alien both to the quantum and classical dynamics of the
system, it is natural to ask whether the approximation makes sense 
quantitatively. 
We show that despite of introducing chaos the inclusion of the
quantum degrees of freedom $(Q_i,P_i)$ gives a better approximation to the
time evolution of the observable $<J_z>(t)$ which we analyse as an example.

In figure 5 we display $<J_z>(t)$ ($\chi=-6.0$ , $J=4$) for the three cases:
The exact calculation, the semiclassical approximation and the classical
result. As can be seen from the figure the semiclassical approximation 
represents an improvement over the classical result. We have checked that
this is always true for short enough times. The validity of the approximation
 is of course sensitive to the value of $\chi$. We next arbitrarily define 
one possible quantitative measure of the accuracy of the approximations.
Consider the expression

\begin{equation}
\Delta_{approx}=\frac{\int |<J_z>_{exact}(t) - <J_z>_{approx}(t)| dt }
{\int |<J_z>_{exact}(t)| dt} 
\end{equation}

In figure 6 we display for three different values of time, the value of
$\Delta_{approx}$ as a function of $1/J$ for both the classical as well as
for the semiclassical approximation.  Notice that the error so defined 
depends
linearly on $1/J$ and the classical calculation lies always above the
semiclassical one. Figure 6 also shows explicitly that in both cases
$\Delta_{approx}$ goes to zero as $J \rightarrow \infty$.
We have also defined a {\it breakdown} time in the following way: 
we fix a maximum
value for the error $\Delta_{approx}^{max}=0.12$ and plot the time $T_b$
when this occurs for several values of $J$ in both cases, the classical and
the semiclassical (gaussian) approximations. Again, according to this 
measure we see that the gaussian approximation is systematically better,
{\it i.e.}, it is valid for longer times (see figure 7).
It is interesting to notice that the form of the curve is the same for
both approximations. The problem of the form of the curve for the breakdown 
time has been the subject of recent investigation \cite{Dittes}.

\newpage

\section{Conclusions:}

\vspace{1.0cm}

In the present contribution we have shown how a mean field expansion in
the sense of nonrelativistic many body theories can be used to obtain
quantum corrections to the classical limit. 
The unitary time evolution of a gaussian state is shown to contain the
 classical limit plus corrections coming from allowing the width of the wave
packet to become an independent variable. We have discussed the connection
between this approach and other approaches in the literature and applied
it to the $SU(2)$ Lipkin model. We have performed a detailed analysis of the
unitary quantum corrections showing that they give rise to chaotic
behavior, which is essentially the mechanism through which the tunneling 
phenomenon can happen in this context. We have also shown that the quantum
corrections systematically improve the results obtained in the classical limit.
The question left unexplored in the present work is the effect of nonunitary
contributions. We believe this is an important next step, {\it i.e.},
including the time evolution of occupation probabilities in the dynamics
 which is rather natural in the present formalism.
It would be interesting to cast these contributions in the form of diffusive 
and dissipative processes. In particular, as can be seen from the equations
of motion, when $\nu_i \neq 0$ , $\nu = \nu(t)$ the centrifugal barrier will
be time dependent, affecting thus in particular the tunneling rates.
Work along these lines is in progress.

Acknowledgment: We are indebted to Prof. A. F. R. de Toledo Piza for many fruitful discussions.

\newpage

{\Large Figure Captions:}

\vspace{1.0cm}
{\bf Figure 1:}

Poincar\'e section on the plane $(q_a,p_a)$ (with $q_b=0$ and $pb>0$) for
the classical dynamics ${\cal{H}}_{cl}(q_i,p_i)$ 
:(a) for $\chi=-0.5<\chi_{crit}$. (b)for $\chi=-6.0>\chi_{crit}$.
See text for details. 

\vspace{0.5cm}
{\bf Figure 2:}

Poincar\'e section on the plane $(q_a,p_a)$ (with $q_b=0$ and $pb>0$) for
the semiclassical dynamics ${\cal{H}}_{sc}(q_i,p_i,Q_i,P_i)$ in the case
$\chi=-6.0>\chi_{crit}$. Initial 
conditions for the widths are chosen in the minimal uncertainty situation
$(Q_i(0)=\sqrt{1/2},P_i=0.0)$. For each initial condition the  energies are 
the same as in figure 1(b) ${\cal{H}}_{sc}={\cal{H}}_{cl}(t=0)$. The values 
of $J$ are:(a) $J=2$, (b) $J=8$ and (c) $J=12$. See text for more details.

\vspace{0.5cm}
{\bf Figure 3:} 
 
Poincar\'e section on the plane $(q_a,p_a)$ (with $q_b=0$ and $pb>0$) 
with $\chi=-6.0>\chi_{crit}$ for a single initial condition near the
separatrix energy ${\cal{H}}_{cl}($S$_-)=-1.0$ in region E$_{min}$,
 $p_a<0.0$: (a) classical evolution with
${\cal{H}}_{cl}=-1.1$ (in arbitrary $\epsilon$ units, $\epsilon=1.0$).
 (b) semiclassical evolution with ${\cal{H}}_{sc}=-1.1$ and $J=9$. 
The initial conditions for the widths are set to the minimal uncertainty

\vspace{0.5cm}
{\bf Figure 4:}  

$\bar{ T }_c / \bar{ T }_p$ evaluated over the first thousand values of $T_c$
as a function of $J$ for various values of ${\cal{H}}_{sc}$ below the 
classical separatrix energy ${\cal{H}}_{cl}($S$_-)=-1.0$.

\vspace{0.5cm}
{\bf Figure 5:}  
Time evolution of $<J_z>(t)$ ($\chi=-6.0$) for the three cases: 
The exact calculation ($J=4$), semiclassical approximation $(J=4)$
 and classical result. Time t is plotted in arbitrary units.

\vspace{0.5cm}
{\bf Figure 6:}  

 Error $\Delta_{approx}$ evaluated at three different times t 
(in arbitrary units) and plotted
 as a function of $1/J$ for both semiclassical and classical approximations.
$\chi=-0.5$.

\vspace{0.5cm}
{\bf Figure 7:}  
Breakdown time $T_b$ (in arbitrary units) for $\Delta_{approx}^{max}=0.1$
plotted as a function of $1/J$ for both semiclassical and classical 
calculations.$\chi=-0.5$.

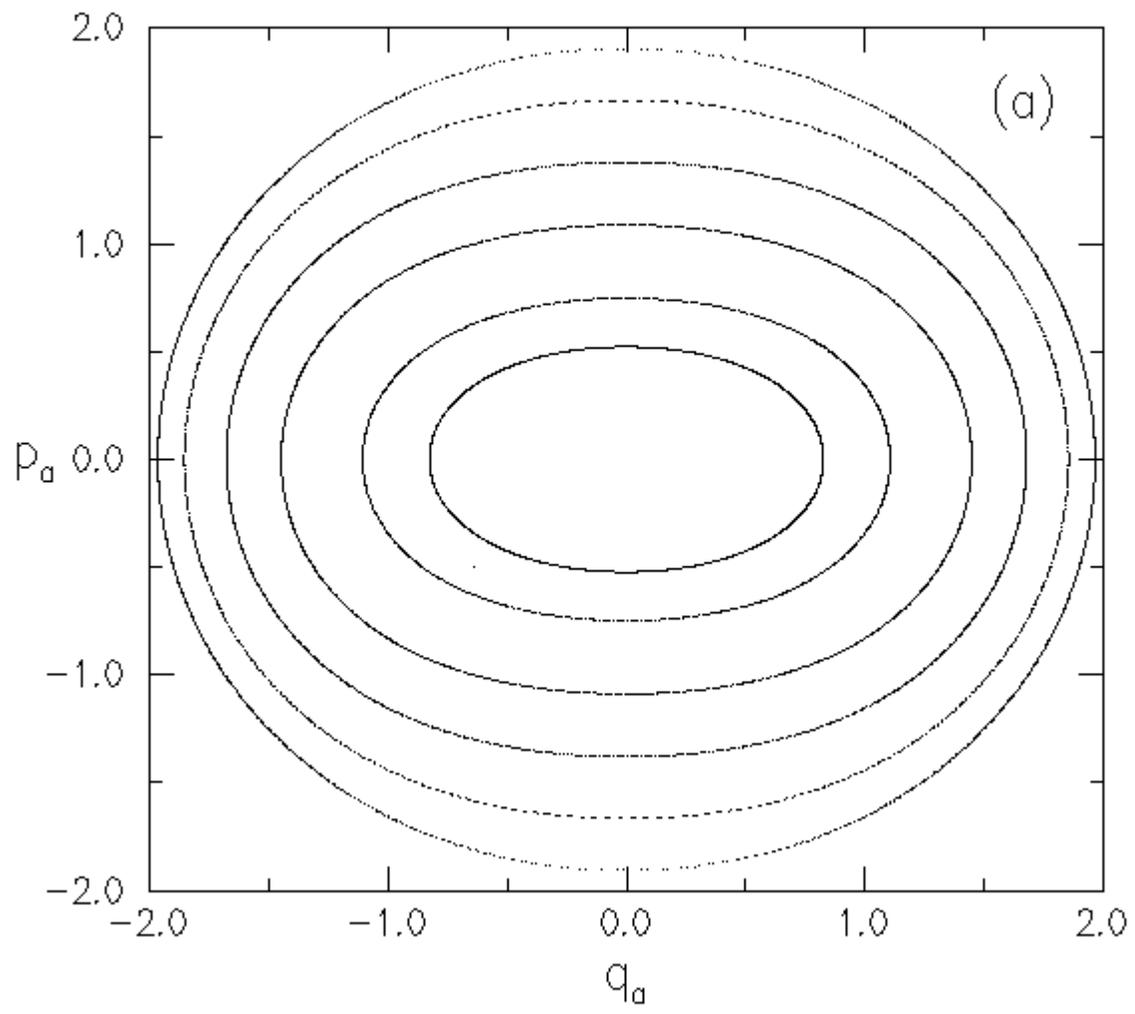
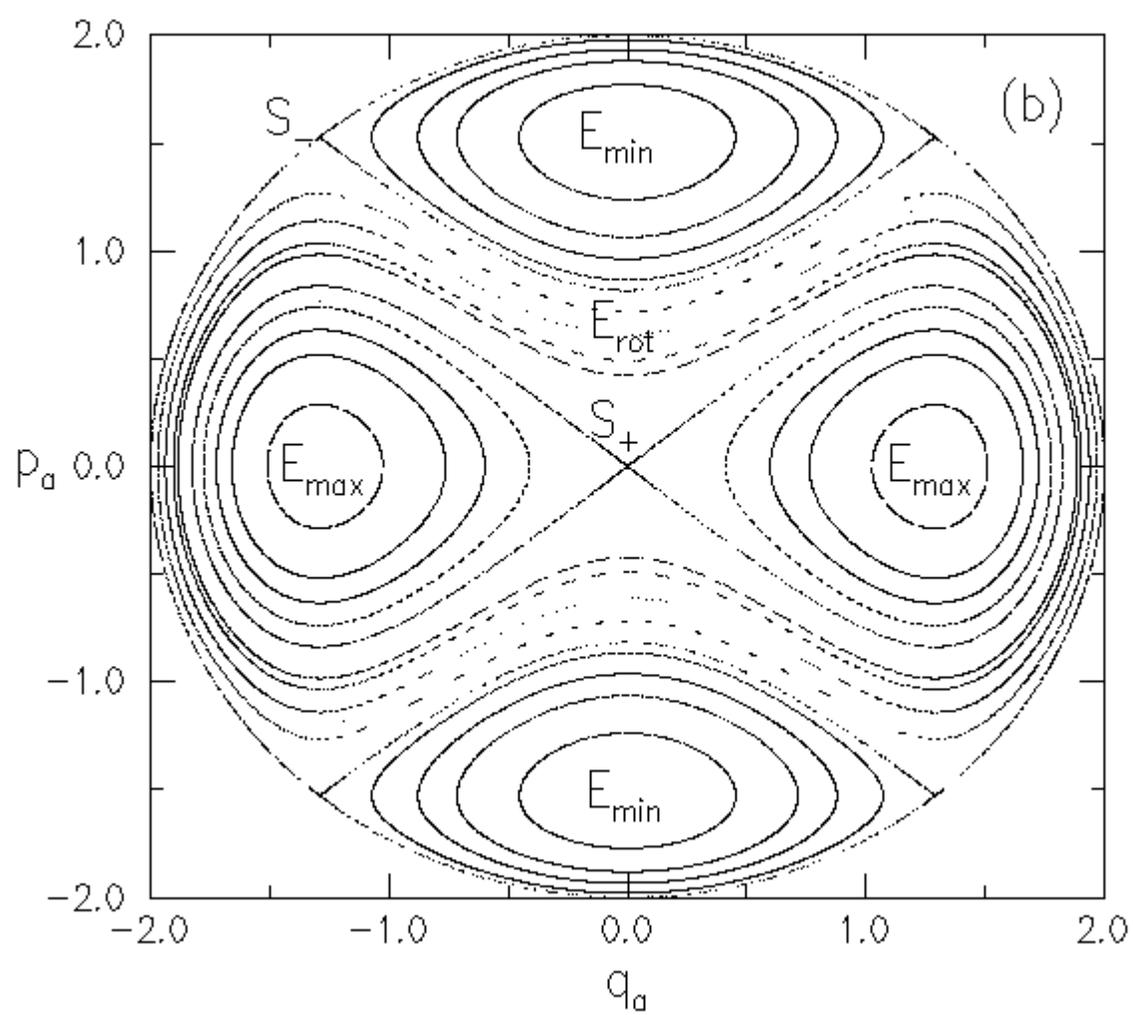
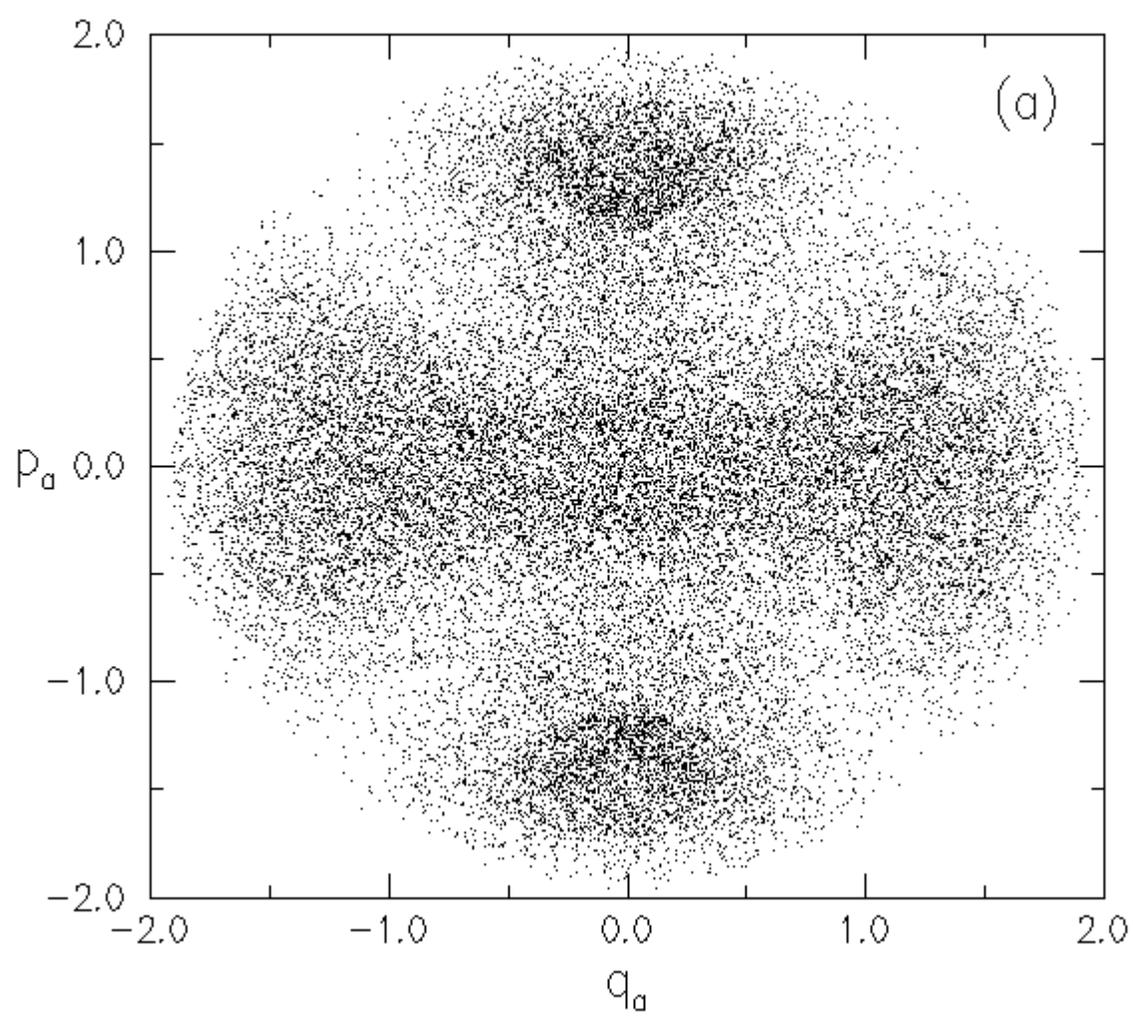
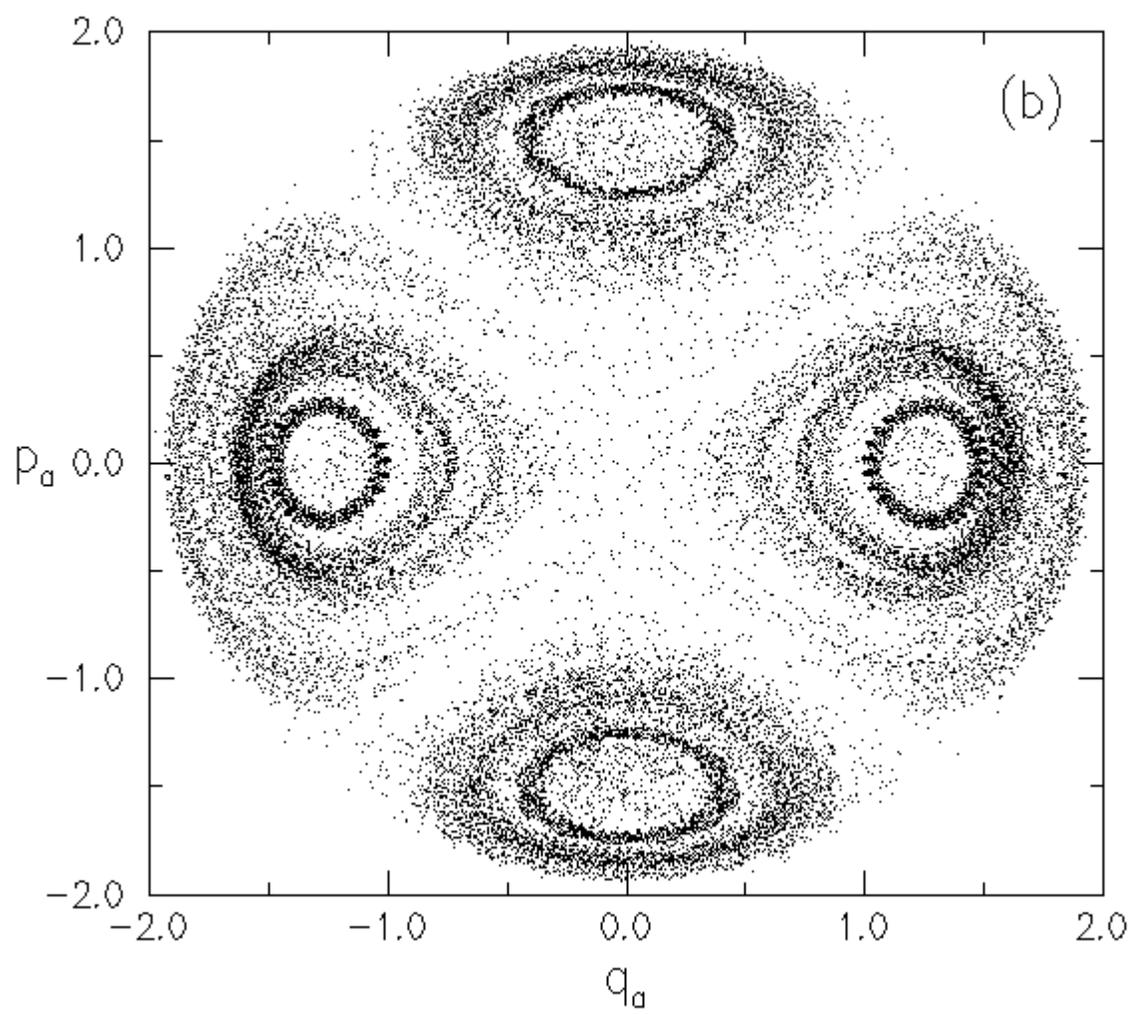
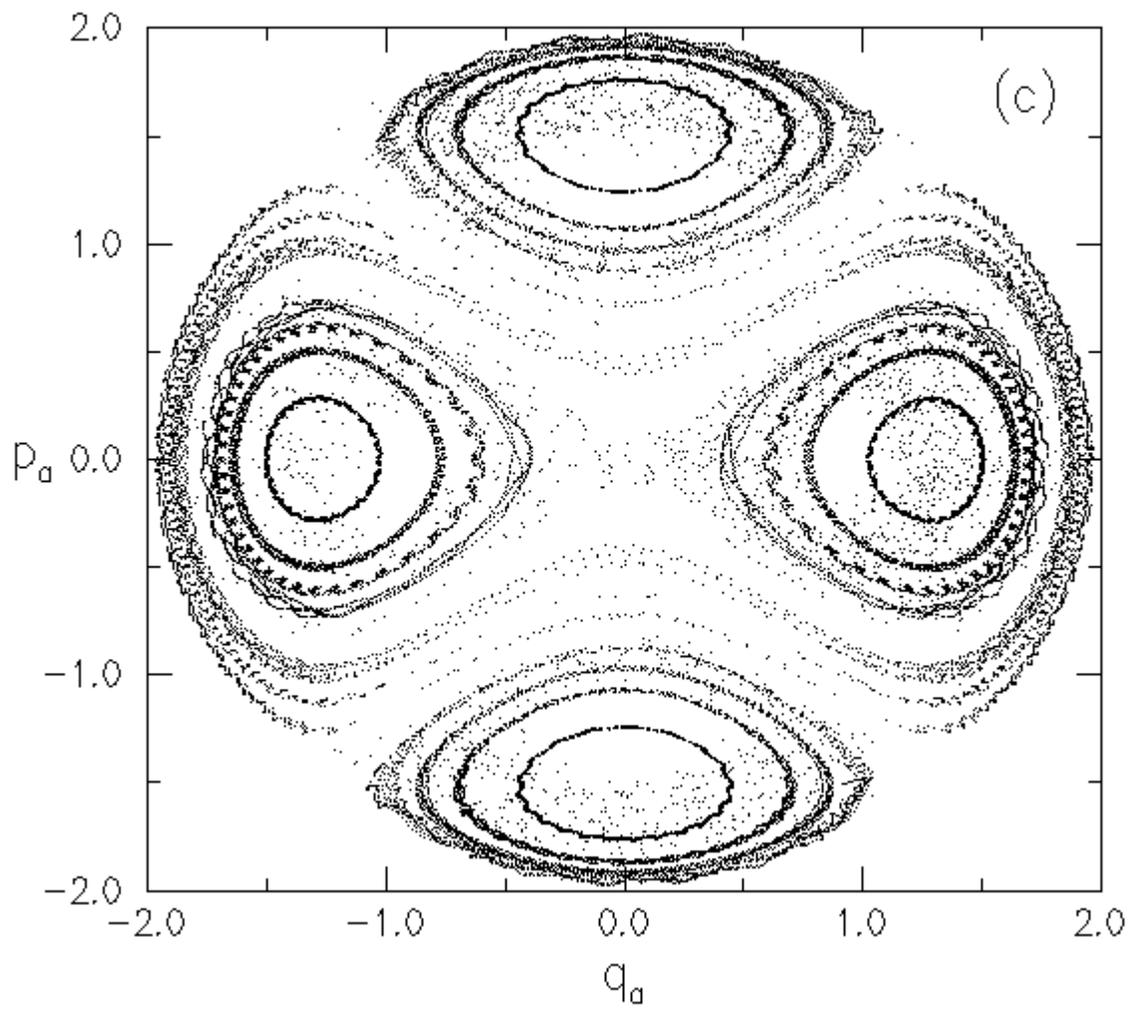
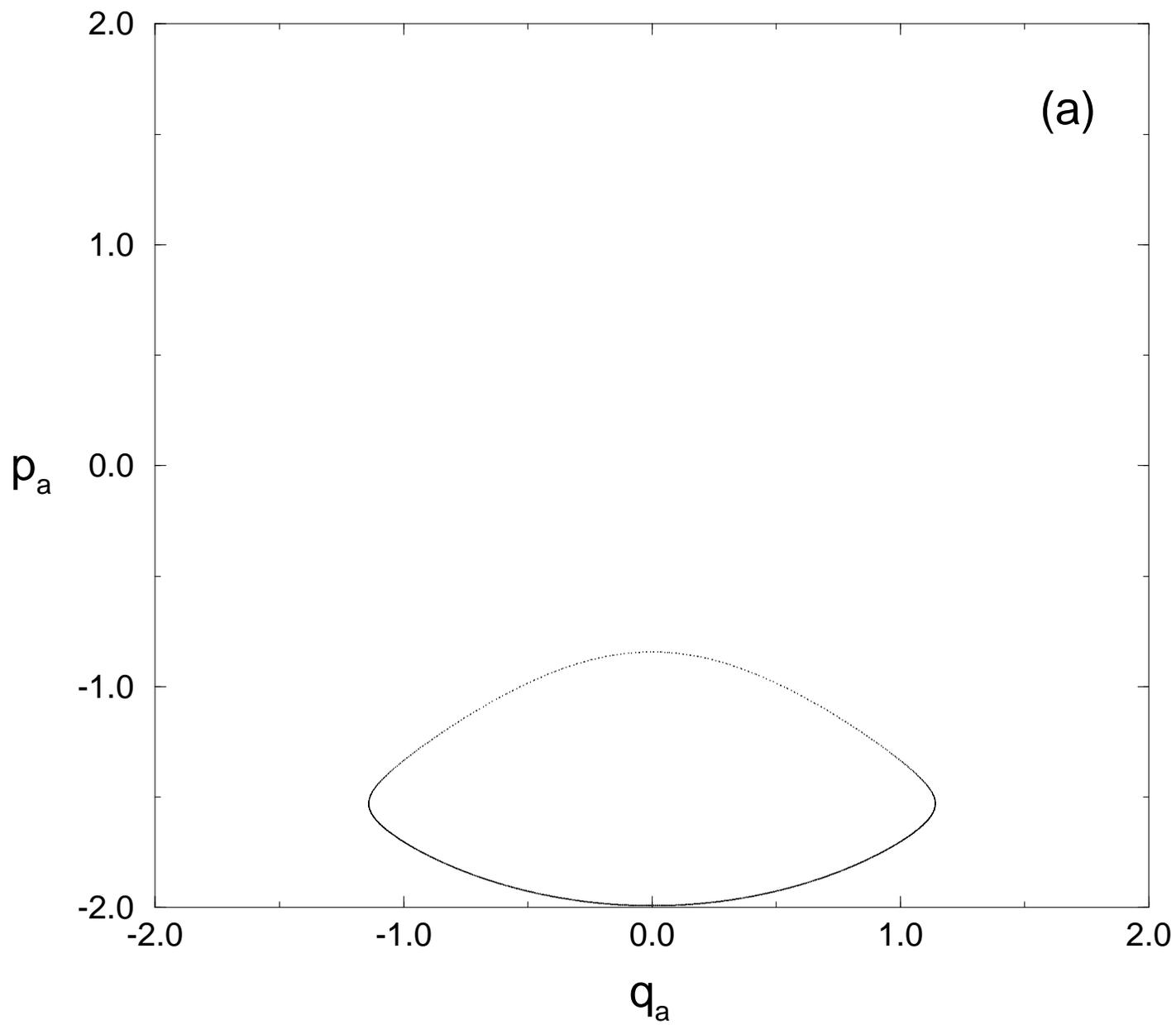
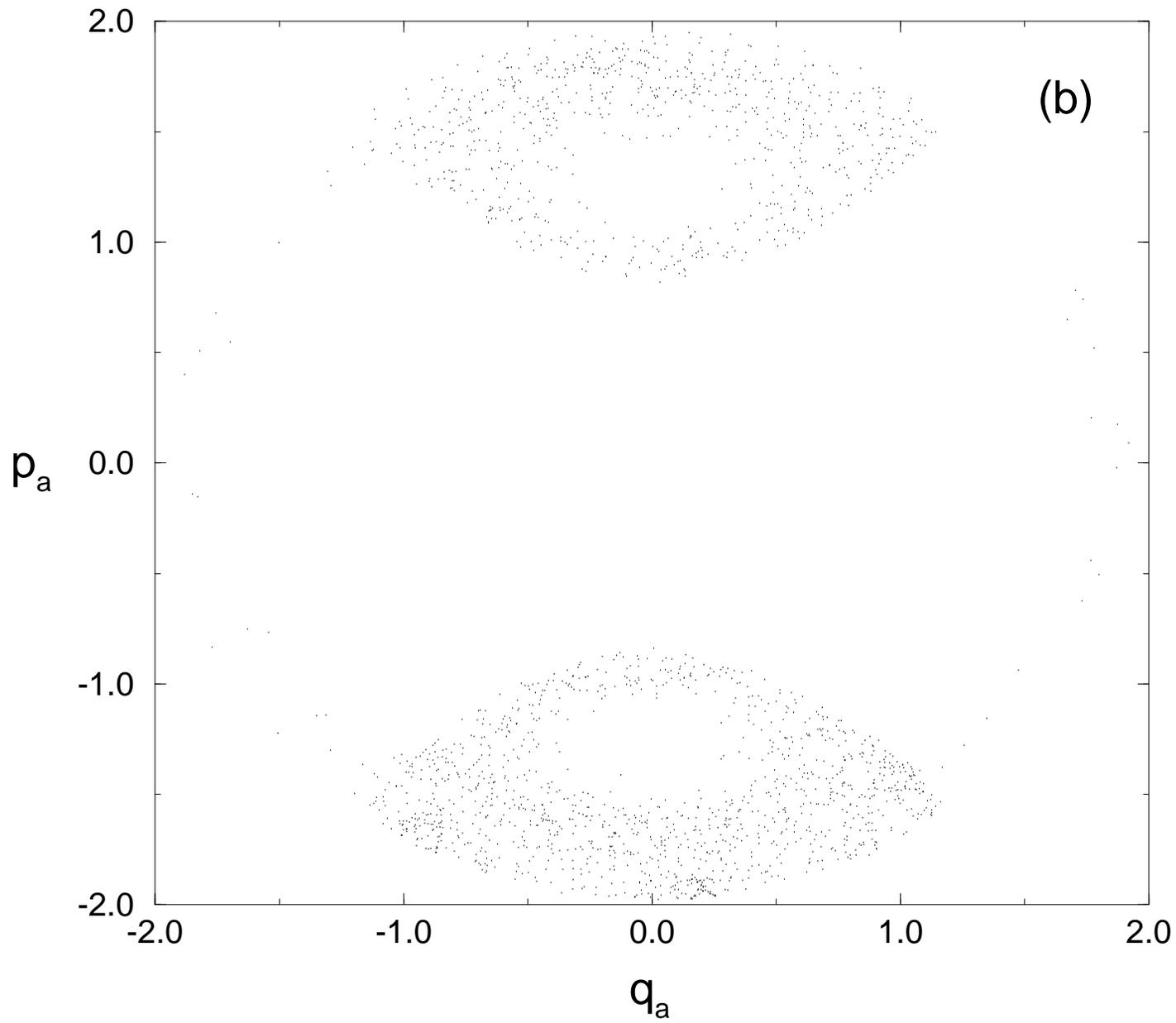
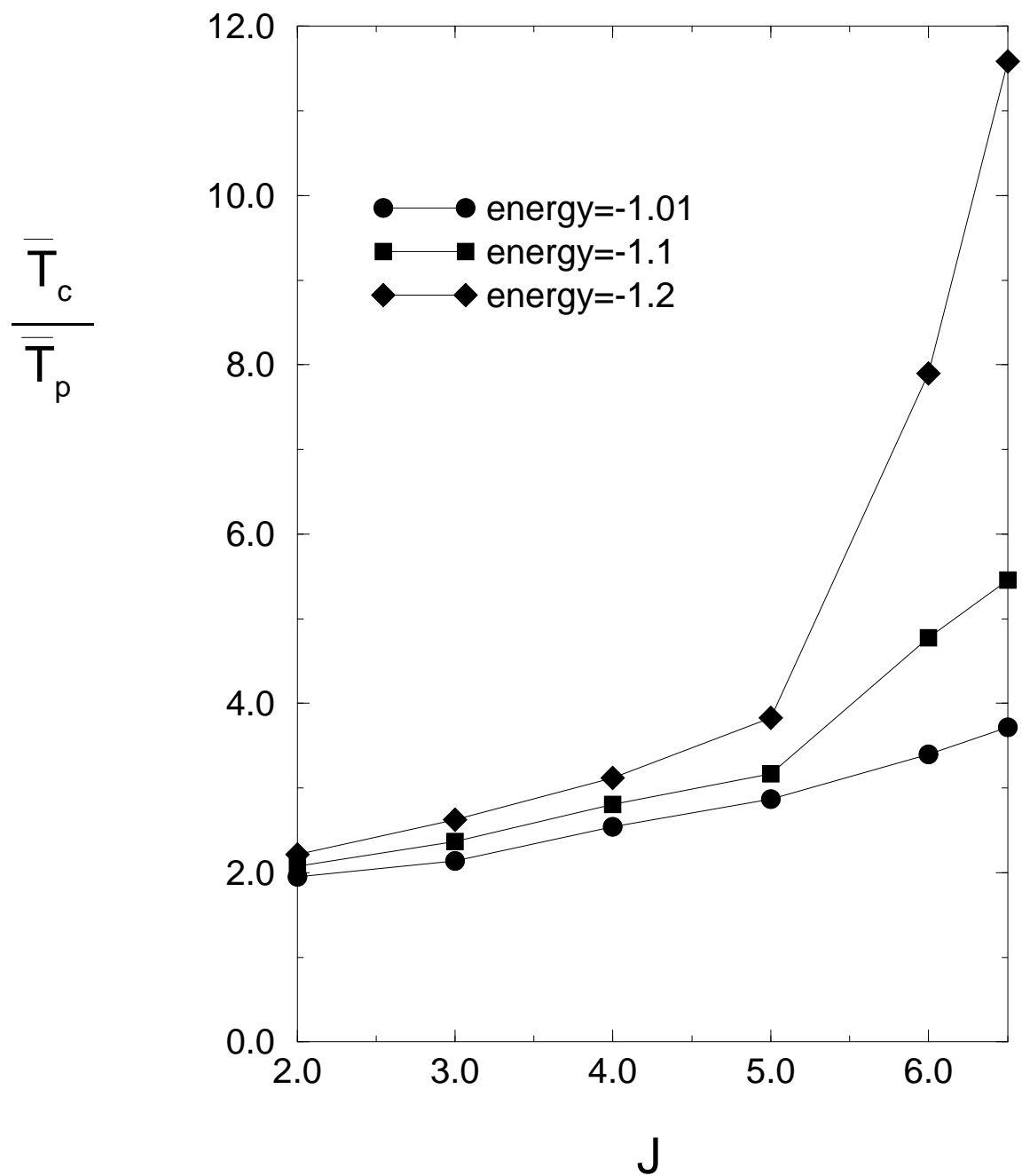
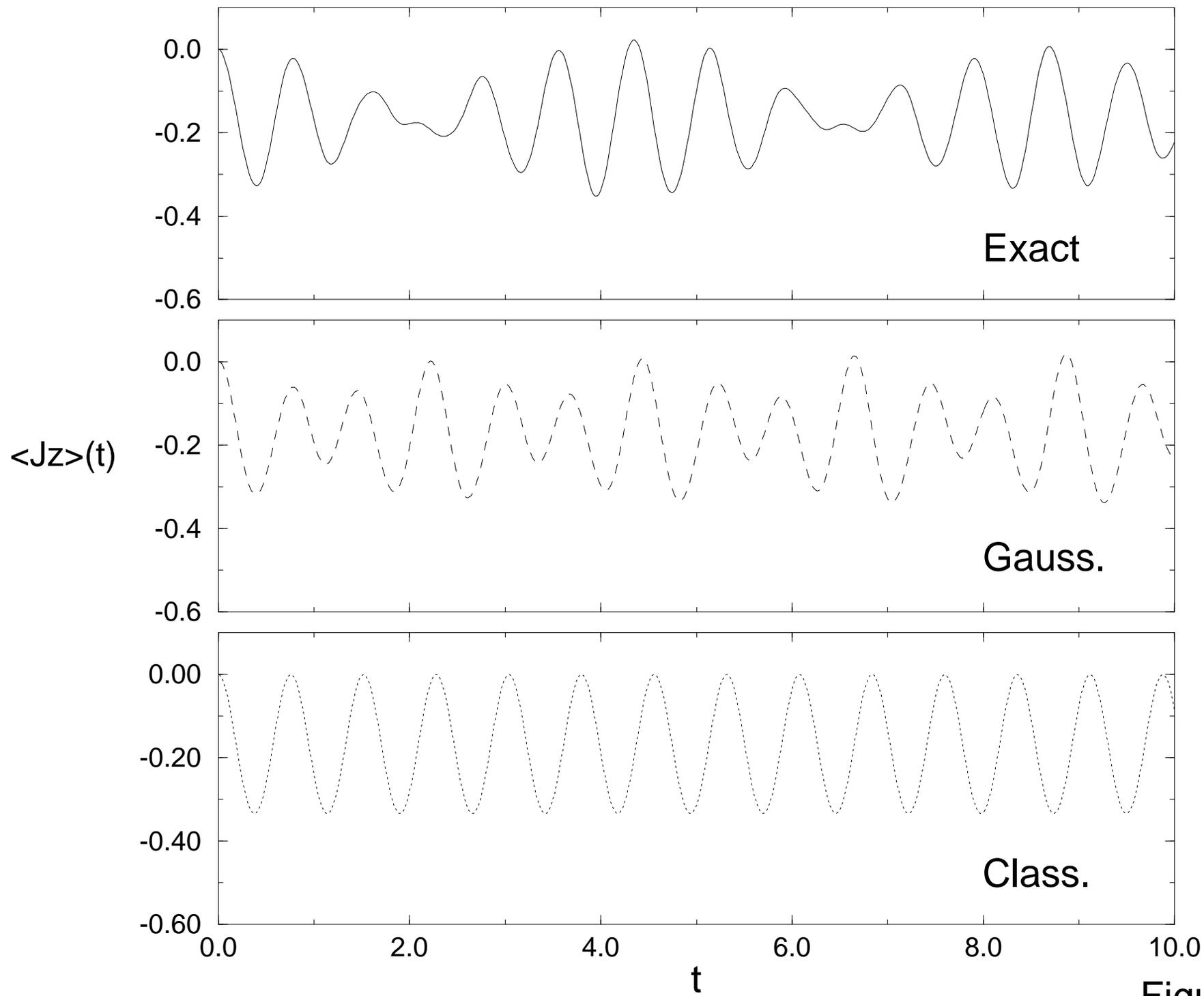
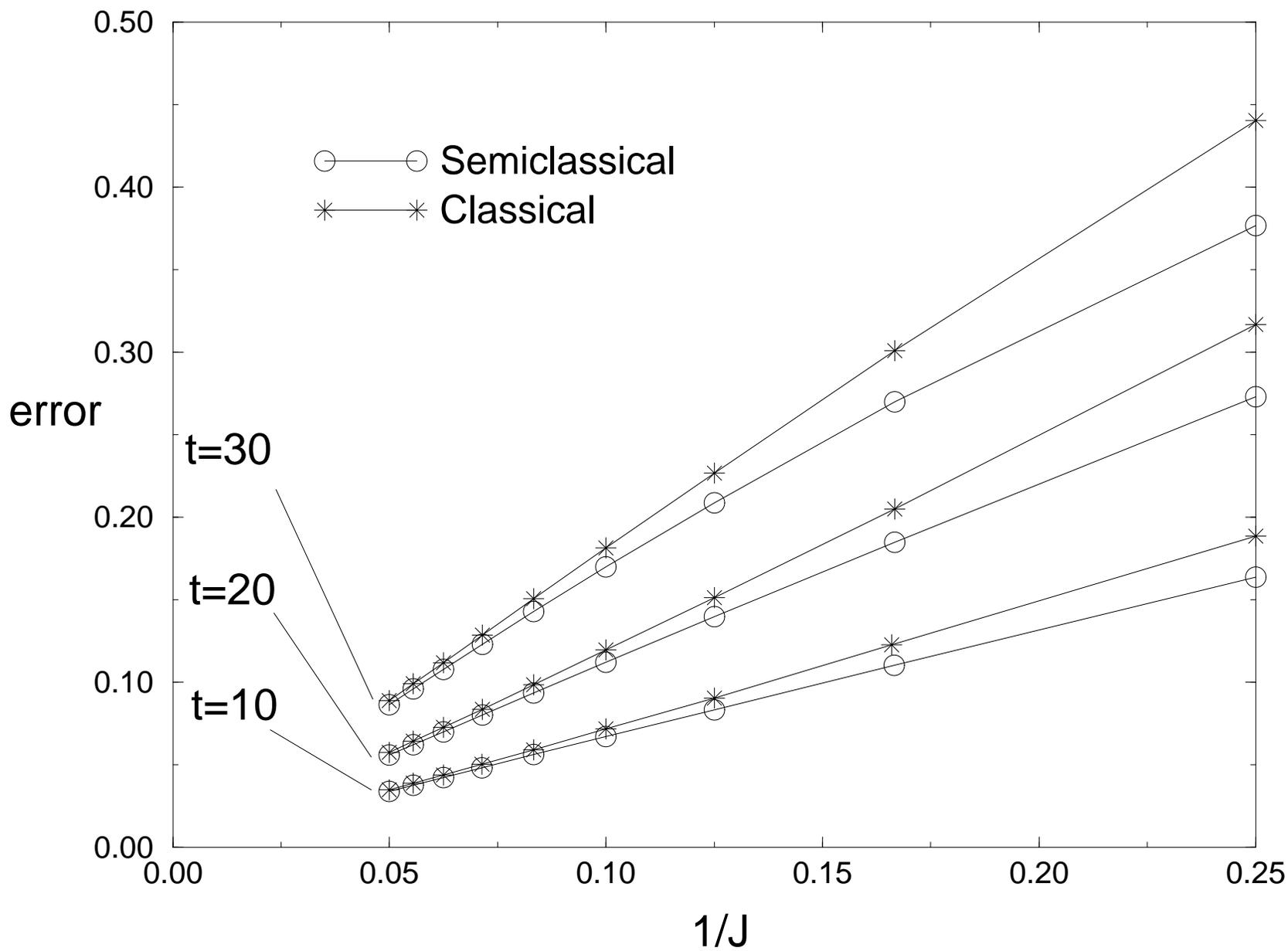
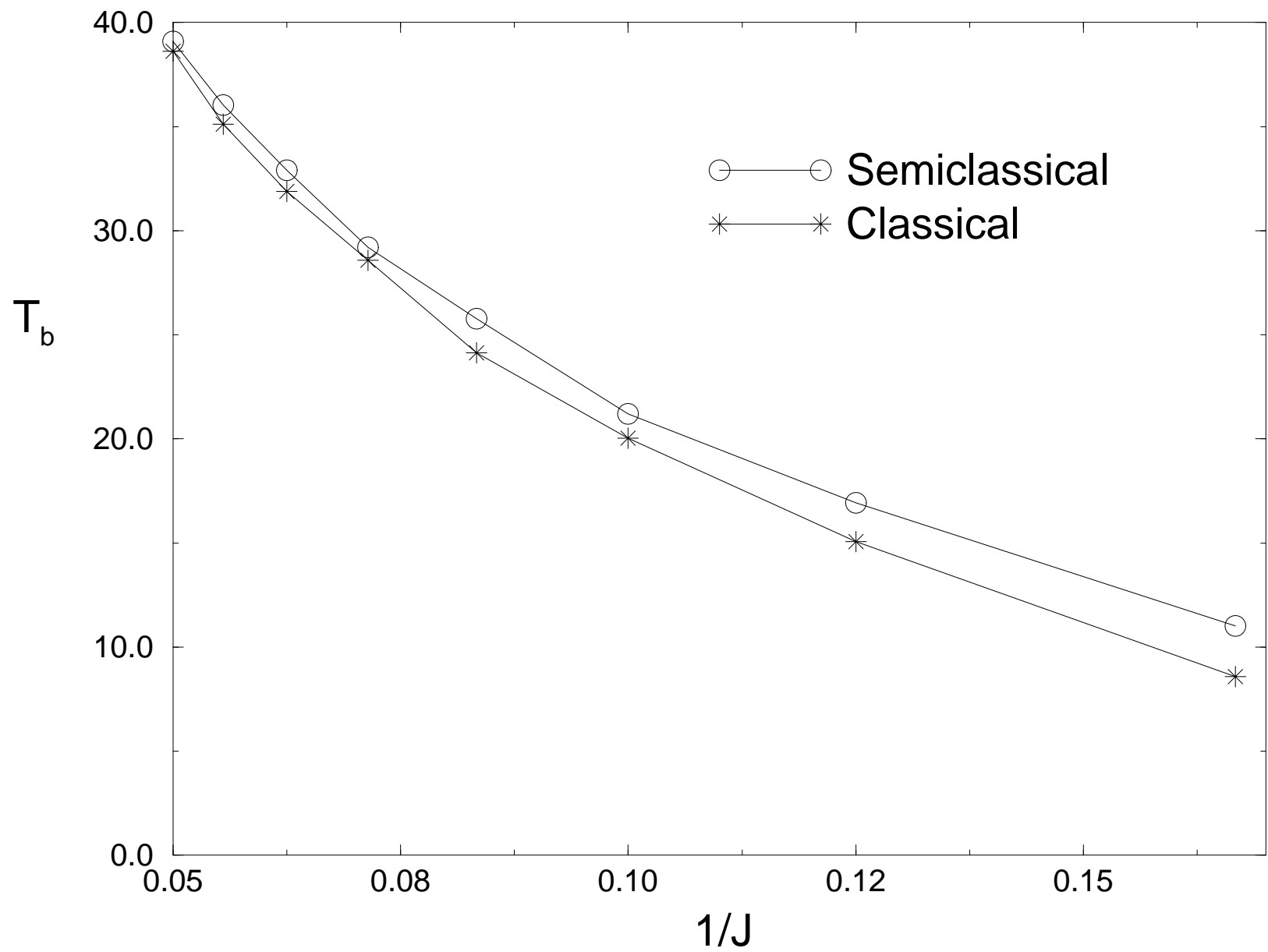
\end{document}